\documentclass[prd,aps,nofootinbib,twocolumn,tightenlines]{revtex4}
\usepackage{amssymb}
\usepackage{amsmath,txfonts}
\usepackage{epsfig}
\allowdisplaybreaks[4]
\usepackage{graphicx}
\usepackage{multirow}
\usepackage{float}
\renewcommand{\arraystretch}{1.8}
\usepackage[a4paper,top=2cm,bottom=2cm,left=2cm,right=2cm,marginparwidth=1.75cm]{geometry}
\usepackage[colorlinks=true,linkcolor=red,citecolor=blue]{hyperref}
\newcommand{\psl}{ p \hspace{-1.8truemm}/ }

\begin{document}
\title{Investigation of $\Lambda_{b}\to \Lambda_{c} \ell^-\overline\nu_\ell$ Decays in Perturbative QCD Approach}

\author{Ying Li$^1$}
\email{liying@ytu.edu.cn}
\author{Jie Chen$^1$}
\author{Ya-Xin Wang$^2$}
\author{Zhi-Tian Zou$^1$}
\email{zouzt@ytu.edu.cn}
\affiliation{
$^1$Department of Physics, Yantai University, Yantai 264005, China\\
$^2$Institute of Particle Physics and Key Laboratory of Quark and Lepton Physics (MOE),
Central China Normal University, Wuhan, Hubei 430079, China}
\vspace{0.2cm}
	
\begin{abstract}
We investigate the semileptonic decays $\Lambda_{b} \to \Lambda_{c} \ell^- \bar{\nu}_\ell$ (with $\ell = e, \mu, \tau$) within the framework of perturbative QCD (pQCD). The six independent $\Lambda_b \to \Lambda_c$ transition form factors are first calculated in the low-$q^2$ region using the $k_T$ factorization approach. These are then extrapolated to the full physical $q^2$ range via the model-independent $z$-expansion, incorporating recent lattice QCD results at high $q^2$. Based on the obtained form factors, we compute the branching fractions of $\Lambda_b \to \Lambda_c \ell^- \bar{\nu}_\ell$ decays. Our prediction for the lepton flavor universality ratio, $\mathcal{R}_{\Lambda_c} = 0.29^{+0.12}_{-0.11}$, is slightly larger than the latest experimental measurement. In addition, we analyze several angular observables, including forward-backward asymmetries, lepton-side convexity parameters, and polarization asymmetries. These results offer valuable theoretical input for current and future experimental investigations of semileptonic heavy-to-heavy baryon transitions.
\end{abstract}
\maketitle
\section{Introduction} \label{sec:1}
Semileptonic decays mediated by the $b\to c\ell^-\bar\nu_\ell$ (with $\ell = e, \mu, \tau$)  transition have shown intriguing deviations from Standard Model (SM) predictions, particularly in the ratios
\begin{eqnarray}\label{RDdef}
\mathcal{R}(D^{(*)}) = \frac{\mathcal{B}(B \to D^{(*)} \tau \bar{\nu}_\tau)}{\mathcal{B}(B \to D^{(*)} \mu \bar{\nu}_\mu)}\,.
\end{eqnarray}
By combining the measurements for both $D$ and $D^*$ final states, the observed deviation from the SM expectation reaches $3.14 \sigma$~\cite{HeavyFlavorAveragingGroupHFLAV:2024ctg}. Precise determination of hadronic matrix elements is essential for accurate predictions of these ratios. In particular, an improved understanding of the $B\to D^{(*)}$ form factors is critical, as they are a major source of the theoretical uncertainties in the SM prediction of $B\to D^{(*)}\ell\bar\nu_\ell$. Moreover, these matrix elements are also central to resolving the tension between inclusive and exclusive determinations of the  Cabibbo-Kobayashi-Maskawa (CKM) matrix element $|V_{cb}|$ ~\cite{Bernlochner:2017jka}.  The observed anomalies have motivated extensive exploration of new physics (NP) beyond the SM, including scenarios involving TeV-scale leptoquarks, exotic gauge bosons, and high-$p_T$ searches at the LHC for potential mediator particles. Further discussions on these topics can be found in the recent comprehensive reviews~\cite{Capdevila:2023yhq, Bifani:2018zmi}.
	
The baryonic decay $\Lambda_b \to \Lambda_c \ell^- \bar{\nu}_\ell$ offers complementary insight into the determination of the CKM matrix element $|V_{cb}|$, which can be extracted from measurements of the differential and total decay rates in $\Lambda_b \to \Lambda_c e^-(\mu^-) \bar{\nu}_\ell$ transitions. Analogous to the ratios defined in $B \to D^{(*)} \ell \bar{\nu}_\ell$ decays, the observable
	\begin{eqnarray} \label{RLambdac}
		\mathcal{R}_{\Lambda_c} = \frac{\mathcal{B}\left(\Lambda_b \to \Lambda_c \tau \bar{\nu}_\tau\right)}{\mathcal{B}\left(\Lambda_b \to \Lambda_c \mu \bar{\nu}_\mu\right)}
	\end{eqnarray}
can be defined, which is particularly sensitive to contributions from physics beyond the SM. The LHCb Collaboration has produced a substantial sample of $\Lambda_b$ baryons, enabling precise studies of their semileptonic decays. Notably, LHCb has already reported a measurement of the semi-tauonic branching fraction and the ratio $\mathcal{R}_{\Lambda_c}$, as presented in Ref.~\cite{LHCb:2022piu}. With further data expected from upcoming runs, a more detailed investigation becomes both timely and promising. In this context, a comprehensive analysis of  $\Lambda_b \to \Lambda_c \ell^- \bar{\nu}_\ell$ decays within the SM is warranted. In addition to $\mathcal{R}_{\Lambda_c}$, a variety of observables can be constructed, many of which exhibit enhanced sensitivity to NP effects. Precise SM predictions for these observables are therefore essential, as they provide a critical baseline for identifying and distinguishing possible deviations induced by different NP scenarios.
	
The semileptonic decay $\Lambda_b \to \Lambda_c \ell^- \bar{\nu}_\ell$ has attracted significant theoretical interest as a complementary channel to mesonic $b \to c\ell\bar{\nu}$ transitions. Within the SM, early studies employed quark models and heavy-quark effective theory (HQET), while recent lattice QCD calculations have provided high-precision predictions for the full set of form factors across the kinematic range. In particular, Detmold, Lehner, and Meinel~\cite{Detmold:2015aaa} calculated the $\Lambda_b \to \Lambda_c$ form factors with percent-level uncertainties, enabling a robust extraction of $|V_{cb}|$ and predictions for observables such as $\mathcal{R}_{\Lambda_c}$. HQET-based analyses incorporating LHCb and lattice inputs have further constrained subleading corrections~\cite{Bernlochner:2018bfn}, enhancing the reliability of SM predictions. Complementary approaches, such as light-cone sum rules~\cite{Duan:2022uzm}, also yield results consistent with lattice data. Theoretical predictions for $\mathcal{R}_{\Lambda_c}$ in the SM lie around $0.33$--$0.34$, in mild tension with recent LHCb measurements~\cite{LHCb:2022piu}, motivating studies of possible new physics (NP) effects. Global analyses have explored various NP scenarios using model-independent effective field theory frameworks~\cite{Mu:2019bin,Fedele:2022iib}, with scalar and tensor operators showing potential to accommodate observed deviations. More recently, angular analyses of the full decay chain $\Lambda_b \to \Lambda_c(\to \Lambda\pi)\tau\bar{\nu}_\tau$ have been performed to identify sensitive observables and constrain NP Wilson coefficients~\cite{Nandi:2024aia}. These efforts underscore the importance of $\Lambda_b$ decays as a testing ground for LFU and as a probe of beyond-the-SM interactions. Continued theoretical refinement and forthcoming experimental data will be essential for resolving current tensions and exploring the full potential of baryonic semileptonic decays.
	
While lattice QCD provides precise form factor predictions in the high-$q^2$ (low recoil) region, its reliability decreases in the low-$q^2$ domain due to discretization and finite-volume effects. On the other hand, the perturbative QCD (pQCD) approach based on $k_T$ factorization is well suited for describing heavy-to-heavy transitions in the large recoil region, where the hard-scattering kernel can be systematically expanded in $\alpha_s$ and nonperturbative effects are encoded in universal hadron light-cone distribution amplitudes (LCDAs)~\cite{Li:2003yj, Keum:2000ph,Lu:2000em}. This framework has been successfully applied to various $B$ meson decays and extended to baryonic transitions, such as $\Lambda_b \to p$ \cite{Shih:1998pb,Han:2022srw}, $\Lambda_b \to \Lambda$ \cite{Yang:2025yaw},  $\Lambda_b \to \Lambda_c$~\cite{Shih:1999yh,Lu:2009cm, Chou:2001bn, Zhang:2022iun}, and $\Xi_b \to \Xi_c$ \cite{Rui:2025iwa}. By calculating the transition form factors within the pQCD approach, one can cross-check and extend lattice QCD results to the low-$q^2$ region, thus providing a more complete description across the entire kinematic range. Furthermore, the $k_T$ factorization formalism includes transverse momentum effects and Sudakov suppression, which regulate endpoint divergences and enhance the theoretical control over power corrections~\cite{Li:1992nu, Kurimoto:2001zj}. A consistent treatment of $\Lambda_b$ and $\Lambda_c$ baryon distribution amplitudes, constrained by heavy quark symmetry and QCD sum rules~\cite{Wang:2009hra, Khodjamirian:2011jp, Wang:2015ndk, Miao:2022bga}, enables reliable numerical predictions. In this work, we shall revisit the $\Lambda_b \to \Lambda_c$ form factors using the pQCD framework, aiming to complement the latest lattice QCD results and to provide robust predictions for observables sensitive to new physics across the full kinematic region accessible to current and future experiments.
	
The content of this paper is as follows. Sec.~\ref{sec:2} is the kinematics and the framework of the pQCD approach used in the calculation of $\Lambda_b \to \Lambda_c$ transition form factors. Sec.~\ref{sec:3} includes wave functions of $\Lambda_b$ and the $\Lambda_c$ baryons. We give the results of $\Lambda_b \to \Lambda_c$ transition form factors in Sec.~~\ref{sec:4}, and the $\Lambda_b \to \Lambda_c \ell^- \bar{\nu}_\ell$ decays in Sec.~\ref{sec:5}, respectively. Finally, we have concluded in Sec.~\ref{sec:6}.
	
\section{Framework} \label{sec:2}
It is convenient to adopt the light-cone coordinate system, in which a four-momentum is written as $p = (p^+, p^-, \mathbf{p}_T)$, with light-cone components defined by $p^{\pm} = \frac{1}{\sqrt{2}}(p^0 \pm p^3)$ and $\mathbf{p}_T = (p^1, p^2)$. The scalar product of two arbitrary four-vectors $A$ and $B$ is then expressed as $A\cdot B=A_{\mu}B^{\mu}= A^+B^- + A^-B^+ - \textbf{A}_{T}\cdot \textbf{B}_{T}$.  For simplicity, we consider the $\Lambda_b$ baryon at rest. We further assume that the final-state $\Lambda_c$ baryon carries a large momentum component along the ``$-$'' light-cone direction. Under these conventions, the momenta of the $\Lambda_b$ and $\Lambda_c$ baryons are given by:
	\begin{eqnarray}\label{eq:pq}
		p=\frac{m_{\Lambda_b}}{\sqrt{2}}\left(1,1,\textbf{0}_{T}\right),\quad
		p^\prime=\frac{m_{\Lambda_{b}}}{\sqrt{2}}\left(\eta_{1},\eta_{2},\textbf{0}_{T}\right).
	\end{eqnarray}
Two components, $\eta_1$ and $\eta_2$, of $p^\prime$ are obtained as
\begin{eqnarray}
	\eta_1 &=& \frac{E - \sqrt{E^2 - 4m_{\Lambda_b}^2 m_{\Lambda_c}^2}}{2m_{\Lambda_b}^2}, \\
	\eta_2 &=& \frac{E + \sqrt{E^2 - 4m_{\Lambda_b}^2 m_{\Lambda_c}^2}}{2m_{\Lambda_b}^2},
\end{eqnarray}
where $E = m_{\Lambda_b}^2 + m_{\Lambda_c}^2 - q^2$, and $q^2 = (p - p^\prime)^2 = m_{\Lambda_b}^2(1 - \eta_1)(1 - \eta_2)$. Here, $m_{\Lambda_b}$ and $m_{\Lambda_c}$ denote the masses of the $\Lambda_b$ and $\Lambda_c$ baryons, respectively. It is evident that $\eta_2$ is of order unity, $\eta_2 \sim \mathcal{O}(1)$, while $\eta_1$ is suppressed as $\eta_1 \sim \mathcal{O}(m_{\Lambda_c}^2/m_{\Lambda_b}^2)$. In the heavy quark limit, the mass difference between the $b$ quark and the $\Lambda_b$ baryon is negligible; thus, we take $m_b \simeq m_{\Lambda_b}$. The masses of the light quarks ($u$, $d$, and $s$) are also neglected in this approximation.
	
In the framework of pQCD, all transverse momenta of the valence quarks are retained. The momenta of the constituent quarks are then assigned as
	\begin{eqnarray}
		k_1=\left(x_1\frac{m_{\Lambda_b}}{\sqrt{2}},\frac{m_{\Lambda_b}}{\sqrt{2}},\textbf{k}_{1T}\right),& &
		k_1^{\prime}=\left(\frac{m_{\Lambda_b}}{\sqrt{2}}\eta_{1},x_{1}^{\prime}\frac{M}{\sqrt{2}}\eta_{2},\textbf{k}_{1T}^{\prime}\right),\nonumber\\
		k_2=\left(x_2\frac{m_{\Lambda_b}}{\sqrt{2}},0,\textbf{k}_{2T}\right),& &
		k_2^{\prime}=\left(0,x_{2}^{\prime}\frac{m_{\Lambda_b}}{\sqrt{2}}\eta_{2},\textbf{k}_{2T}^{\prime}\right),\nonumber\\
		k_3=\left(x_3\frac{m_{\Lambda_b}}{\sqrt{2}},0,\textbf{k}_{3T}\right),& &
		k_3^{\prime}=\left(0,x_{3}^{\prime}\frac{m_{\Lambda_b}}{\sqrt{2}}\eta_{2},\textbf{k}_{3T}^{\prime}\right).
	\end{eqnarray}
Here, the $b$ and $c$ quarks are treated as massive and carry momenta $k_1$ and $k^\prime_1$, respectively.  The momenta $k_{2}\left(k_{3}\right)$ and $k_{2}^{\prime}$ ($k_{3}^{\prime}$) correspond to the spectator $u$ ($d$) quarks in the $\Lambda_b$ and $\Lambda_c$ baryons, respectively. The $x_{i}$ and $x_{i}^{\prime}$ denote the longitudinal momentum fractions of the valence quarks,and  $ \textbf{k}_{iT}, \textbf{k}_{iT}^{\prime}$represent their transverse momenta. These momenta satisfy the following conservation relations:
	\begin{eqnarray}
		x_1+x_2+x_3=1,\quad \textbf{k}_{1T}+\textbf{k}_{2T}+\textbf{k}_{3T}=0,
	\end{eqnarray}
and the same conditions hold for the primed quantities.
	
Factorization is a central concept in applying pQCD to hard exclusive processes, as it allows the separation of long-distance (nonperturbative) dynamics from short-distance (perturbative) dynamics. The hadronic matrix elements for the $\Lambda_{b} \to \Lambda_{c}$ transition can be expressed as
	\begin{eqnarray}
		\mathcal{M}\sim \Psi_{\Lambda_{b}}\left(x_{i},b_{i},\mu\right)\otimes H\left(x_{i},b_{i},x_{i}^{\prime},b_{i}^{\prime},\mu\right)\otimes \Psi_{\Lambda_{c}}\left(x_{i}^{\prime},b_{i}^{\prime},\mu\right),
	\end{eqnarray}
where $H$ denotes the hard scattering kernel, and $\Psi_{\Lambda_{b}}$ and $\Psi_{\Lambda_{c}}$ are the wave functions of the $\Lambda_{b}$ and $\Lambda_{c}$ baryons, respectively. Both the wave functions and the hard kernel $H$ depend on the factorization scale $\mu$, which serves as the boundary separating soft and hard dynamics. Above this scale, the dynamics are governed by perturbative (short-distance) interactions; below it, the dynamics are non-perturbative and are absorbed into the hadron wave functions. The factorization scale is typically chosen to be equal to the renormalization scale for convenience. In practical calculations, it is advantageous to work in transverse coordinate space (commonly referred to as $b$-space) rather than in transverse momentum space ($\mathbf{k}_T$-space). Accordingly, a Fourier transformation is performed to convert the wave functions and the hard amplitude into $b$-space. The variable $1/b$ naturally emerges as a typical scale separating hard and soft regions. For scales $\mu > 1/b$, the interactions are dominated by short-distance QCD dynamics and are computed perturbatively. Conversely, for $\mu < 1/b$, soft contributions dominate and are included in the non-perturbative wave functions. This treatment ensures a consistent separation of perturbative and non-perturbative effects within the $k_T$ factorization formalism.

Higher-order radiative corrections to both the wave functions and the hard scattering amplitudes give rise to large double logarithmic terms of the form $\alpha_s \ln^2(Q b)$, which originate from the overlap of collinear and soft divergences. To ensure the validity of the perturbative expansion, such large logarithms must be systematically resummed. Resummation techniques have been developed for this purpose, leading to the appearance of a Sudakov exponential factor, $\exp[-s(Q, b)]$, which suppresses contributions from large transverse separations. This exponential falls rapidly with increasing $b$ and vanishes as $b \to 1/\Lambda_{\rm QCD}$. As a result, the full hadronic wave functions $\Psi_{\Lambda_b}(x,{\bf{b}},p,\mu)$ and $\Psi_{\Lambda_c} (x^{\prime},{\bf{b}}^{\prime},p^{\prime},\mu)$ can be factorized into products of the Sudakov exponentials and reduced (non-perturbative) wave functions, denoted by $\Phi_{\Lambda_b}(x,{\bf{b}},p,\mu)$ and $\Phi_{\Lambda_c}(x^{\prime},{\bf{b}}^{\prime},p^{\prime},\mu)$
	\begin{widetext}
		\begin{eqnarray}
			\Psi_{\Lambda_b}(x,{\bf{b}},p,\mu)=\exp\left[-\sum_{i=2}^{3}
			s(w,k_i^{+})\right]\Phi_{\Lambda_b}(x,{\bf{b}},p,\mu)\;
			,\nonumber \\
			\Psi_{\Lambda_c}
			(x^{\prime},{\bf{b}}^{\prime},p^{\prime},\mu)=\exp\left[-\sum_{i=1}^{3}
			s(w^{\prime},k_i^{\prime -})\right] \Phi_{\Lambda_c}
			(x^{\prime},{\bf{b}}^{\prime},p^{\prime},\mu). \label{sp}
		\end{eqnarray}
The Sudakov exponent function $s(b, Q)$ is given by~\cite{Botts:1989kf}:
		\begin{eqnarray}
			s(b,Q)&=&~~\frac{A^{(1)}}{2\beta_{1}}\hat{q}\ln\left(\frac{\hat{q}}
			{\hat{b}}\right)-
			\frac{A^{(1)}}{2\beta_{1}}\left(\hat{q}-\hat{b}\right)+
			\frac{A^{(2)}}{4\beta_{1}^{2}}\left(\frac{\hat{q}}{\hat{b}}-1\right)
			-\left[\frac{A^{(2)}}{4\beta_{1}^{2}}-\frac{A^{(1)}}{4\beta_{1}}
			\ln\left(\frac{e^{2\gamma_E-1}}{2}\right)\right]
			\ln\left(\frac{\hat{q}}{\hat{b}}\right)
			\nonumber \\
			&&+\frac{A^{(1)}\beta_{2}}{4\beta_{1}^{3}}\hat{q}\left[
			\frac{\ln(2\hat{q})+1}{\hat{q}}-\frac{\ln(2\hat{b})+1}{\hat{b}}\right]
			+\frac{A^{(1)}\beta_{2}}{8\beta_{1}^{3}}\left[
			\ln^{2}(2\hat{q})-\ln^{2}(2\hat{b})\right],
		\end{eqnarray}
	\end{widetext}
	with
	\begin{eqnarray}
		\hat q\equiv \mbox{ln}[Q/(\sqrt 2\Lambda)],~~~ \hat b\equiv
		\mbox{ln}[1/(b\Lambda)].
	\end{eqnarray}
	The coefficients $A^{(i)}$ and $\beta_i$ are
	\begin{eqnarray}
		\beta_1=\frac{33-2n_f}{12},~~\beta_2=\frac{153-19n_f}{24},\nonumber\\
		A^{(1)}=\frac{4}{3},~~A^{(2)}=\frac{67}{9}
		-\frac{\pi^2}{3}-\frac{10}{27}n_f+\frac{8}{3}\beta_1\mbox{ln}(\frac{1}{2}e^{\gamma_E}),
	\end{eqnarray}
where $n_f$ is the number of active quark flavors and $\gamma_E$ is the Euler–Mascheroni constant. In our calculations, we employ the one-loop running coupling constant, which corresponds to retaining only the first four terms in the expression for the Sudakov exponent $s(b, Q)$.

Apart from the double logarithms arising from the inclusion of transverse momentum, large single logarithms originating from ultraviolet divergences also appear in the radiative corrections to both the hadronic wave functions and the hard kernels. These single logarithms can be resummed using the renormalization group (RG) method. As a result, we obtain:
\begin{widetext}
	\begin{eqnarray}
		\Phi_{\Lambda_b}(x,{\bf{b}},p,\mu)&=&\exp\left[-{8 \over
			3}\int_{\kappa w}^{\mu} \frac{d\bar{\mu}}{\bar{\mu}}\gamma
		_q(\alpha_s(\bar{\mu}))\right] \Phi_{\Lambda_b}(x,{\bf{b}},p,w), \nonumber \\
		\Phi_{\Lambda_c}(x^{\prime},{\bf{b}^{\prime}},p^{\prime},\mu)&=&\exp\left[-3\int_{\kappa
			w^{\prime}}^{\mu} \frac{d\bar{\mu}}{\bar{\mu}}\gamma
		_q(\alpha_s(\bar{\mu}))\right]  \Phi_{\Lambda_c}(x^{\prime},{\bf{b}^{\prime}},p^{\prime},w^{\prime}), \nonumber \\
		H(x,x^{\prime},{\bf{b}},{\bf{b}}^{\prime},\mu)
		&=&\exp\left[-{17 \over
			3}\,\int_{\mu}^{t}\frac{d\bar{\mu}}{\bar{\mu}}
		\gamma_q(\alpha_s(\bar{\mu}))\right] H(x,x^{\prime},{\bf{b}},{\bf{b}}^{\prime},t)\;,
	\end{eqnarray}
\end{widetext}
where the quark anomalous dimension in the axial gauge is given by $\gamma_q = -\alpha_s/\pi$. The factorization scales $w$ and $w^{\prime}$ characterize the inverse of typical transverse separations among the three valence quarks in the initial and final baryons, respectively, and are defined as
\begin{eqnarray}
	w={\rm min} ({1 \over b_1}, {1 \over b_2}, {1\over b_3}), \qquad
	w^{\prime}={\rm min} ({1 \over b_1^{\prime}}, {1 \over
		b_2^{\prime}}, {1\over b_3^{\prime}}), \label{factorizable scale}
\end{eqnarray}
where
\begin{eqnarray}
	b_1 = |{\bf b}_2 - {\bf b}_3|, \qquad b_1^{\prime} = |{\bf b}_2^{\prime} - {\bf b}_3^{\prime}|,
\end{eqnarray}
and the remaining $b_i$ and $b_i^{\prime}$ are obtained by cyclic permutations. The parameter $\kappa$ is introduced to characterize different schemes for distributing radiative corrections between the perturbative Sudakov factor and the non-perturbative wave function. The value $\kappa=1.14$ is adopted following the fit to the proton form factor~\cite{Kundu:1998gv}.

It should be noted that radiative corrections to the hard amplitudes can generate large double logarithms of the form $\alpha_s \ln^2 x$, which must also be resummed to all orders. The resulting threshold resummation function $S_t(x)$ vanishes rapidly in the end-point regions, $x \to 0$ and $x \to 1$, thereby suppressing contributions from these regions in processes such as $B$ meson decays to light mesons. In contrast, this suppression is less significant in $B \to D$ transitions. Since the $\Lambda_b \to \Lambda_c$ transition also falls into the category of heavy-to-heavy processes, we adopt the simplification $S_t(x) = 1$ in our analysis.

Finally, the matrix element for the $\Lambda_{b} \to \Lambda_{c}$ transition can be expressed as
	\begin{eqnarray}
		\mathcal{M}&\sim &\Phi_{\Lambda_{b}}\left(x_{i},b_{i},\omega\right)\otimes H\left(x_{i},b_{i},x_{i}^{\prime},b_{i}^{\prime},t\right)
		\otimes \Phi_{\Lambda_{c}}\left(x_{i}^{\prime},b_{i}^{\prime},\omega^\prime\right)\nonumber\\
		&\otimes&\exp[-S],\label{convolution}
	\end{eqnarray}
where the explicit form of the Sudakov factor $S$ is given by~\cite{Lu:2009cm}
	\begin{eqnarray}
		S&=&\sum_{l=2,3}s(w,k^+_l)+ \sum_{l=1,2,3}s(w^\prime,k^{\prime-}_l)\nonumber\\
		&+&\frac{8}{3}\int^{t}_{\kappa w}\frac{d\bar \mu}{\bar \mu}\gamma_q(\alpha_s(\bar \mu))+3\int^t_{\kappa w'}\frac{d\bar \mu}{\bar \mu}\gamma_q(\alpha_s(\bar \mu)).
	\end{eqnarray}
	
\section{The Wave Functions} \label{sec:3}
The wave functions of both the initial and final states are among the most crucial input parameters in the pQCD approach. Both the $\Lambda_b$ and $\Lambda_c$ baryons are composed of a heavy quark and two light quarks. In the heavy quark limit, the dynamics of the heavy quark can be factorized from the light degrees of freedom. Assuming that the orbital and spin degrees of freedom of the light-quark system decouple, the leading-twist light-cone distribution amplitude (LCDA) of the $\Lambda_b$ baryon can be written as~\cite{Loinaz:1995wz},
	\begin{widetext}
		\begin{eqnarray}			\Phi_{\Lambda_b,\alpha\beta\gamma}&=&\frac{1}{2\sqrt{2}N_c}\int \prod_{l=2}^3\frac{dy_l^-d\textbf{y}_l}{(2\pi)^3}
			e^{ik_l\cdot y_l}\epsilon^{abc}\langle 0|T[b^a_\alpha(y_1)u^b_\beta(y_2)d^c_\gamma(0)]|u_{\Lambda_b}(p,s)\rangle
			\nonumber\\
			&=&\frac{f_{\Lambda_b}}{8\sqrt{2}N_c}[(\psl+m_{\Lambda_b})\gamma_5 C]_{\beta\gamma}[u_{\Lambda_b}(p,s)]_\alpha
			\phi_{\Lambda_b}(k_i,\mu),
		\end{eqnarray}
	\end{widetext}
where $u_{\Lambda_{b}}$ denotes the heavy baryon spinor, $a$, $b$, and $c$ are color indices, and $\alpha$, $\beta$, and $\gamma$ are spinor indices. $N_c$ is the number of colors, and $C$ is the charge conjugation matrix. The parameter $f_{\Lambda_b} = 2.71 \times 10^{-3}~\mathrm{GeV}^{2}$~\cite{Shih:1999yh} is the normalization constant. Under similar assumptions, the distribution amplitude of the $\Lambda_{c}$ baryon, $\Phi_{\Lambda_c, \alpha^{\prime} \beta^{\prime}\gamma^{\prime}}$, can be written as
	\begin{widetext}
		\begin{eqnarray}
			\Phi_{\Lambda_c,\alpha^{\prime}\beta^{\prime}\gamma^{\prime}}&=&\frac{1}{2\sqrt{2}N_c}\int\prod_{l=2}^3\frac{dy_l^{\prime}d\textbf{y}^{\prime}_l}{(2\pi)^3}
			e^{ik_l^{\prime}\cdot y^{\prime}_l}\epsilon^{abc}\langle0|T[c^a_{\alpha^\prime}(y^{\prime}_1)u_{\beta^\prime}^b(y^{\prime}_2)d^c_{\gamma^\prime}(0)]|u_{\Lambda_c}(p',s')\rangle\nonumber\\
			&=&\frac{f_{\Lambda_c}}{8\sqrt{2}N_c}[(\psl^\prime+m_{\Lambda_c})\gamma_5 C]_{\beta^{\prime}\gamma^{\prime}}[u_{\Lambda_c}(p',s')]_{\alpha^{\prime}}
			\phi_{\Lambda_c}(k_i^{\prime},\mu).
		\end{eqnarray}
	\end{widetext}
In the heavy quark limit, the approximate relation$f_{\Lambda_b}m_{\Lambda_b}=f_{\Lambda_c}m_{\Lambda_c}$~\cite{Shih:1999yh} is satisfied. Therefore, we take $f_{\Lambda_c} = 6.66 \times 10^{-3}~\mathrm{GeV}^{2}$.	
	
The phenomenological model for the distribution amplitude of the $\Lambda_b$ baryon employed in this work is given by~\cite{Schlumpf:1992ce}
\begin{align} \label{lcdasofb}
	&\phi_{\Lambda_b}(x_1,x_2,x_3)\nonumber\\
	&=N_{\Lambda_b}x_1x_2x_3\exp\left[-\frac{1}{2\beta_b^2}\left(\frac{m_b^2}{x_1}+\frac{m_u^2}{x_2}+\frac{m_d^2}{x_3}\right)\right],
\end{align}
where $\beta_b$ is the shape parameter and $m_q$ ($q = u, d, b$) denotes the masses of the constituent quarks in the baryon. The normalization constant $N_{\Lambda_b}$ is determined by
\begin{eqnarray}
	\int dx_1\,dx_2\,dx_3\, \delta(x_1 + x_2 + x_3 - 1)\,\phi_{\Lambda_b}(x_1,x_2,x_3) = 1.
\end{eqnarray}
It is evident that the function $\phi_{\Lambda_b}(x_1,x_2,x_3)$ is symmetric under the exchange of the two light quarks. For the LCDA of the $\Lambda_c$ baryon, we adopt the same form, since the masses of the bottom and charm quarks are both significantly larger than those of the light quarks.  The above forms have been adopted in various phenomenological studies~\cite{Shih:1999yh, Chou:2001bn, He:2006vz}.

Recent developments have significantly advanced our understanding of the LCDAs of heavy baryons $\Lambda_b$ and $\Lambda_c$. At leading twist, the $\Lambda_b$ LCDA has been studied within the framework of HQET, under the assumption that the light diquark system forms a spin-zero configuration. The RG evolution of the leading-twist LCDA has been derived using light-cone operator techniques~\cite{Ball:2008fw, Bell:2013tfa}. Higher-twist three-quark LCDAs have also been systematically classified and evaluated via QCD sum rules~\cite{Khodjamirian:2011jp, Ali:2012zza}, revealing that the distributions are sharply peaked when the heavy quark carries the majority of the baryon's momentum. Similar studies have been extended to the $\Lambda_c$ system under analogous assumptions~\cite{Wang:2009hra, Li:2021qod}, indicating that pQCD factorization can remain valid in specific heavy-to-heavy kinematic regimes. Recent progress also includes next-to-leading-order corrections to the evolution equations and normalization conditions, which have improved the theoretical precision of these frameworks. Concurrently, lattice QCD has begun providing first-principles estimates for low-order moments of the $\Lambda_b$ LCDA~\cite{Detmold:2013nia, Meinel:2016dqj}, offering increasingly stringent constraints on phenomenological models. However, as the precision of LCDA determinations improves, the introduction of additional shape parameters can paradoxically amplify theoretical uncertainties. To avoid this, we still adopt the form in Eq.~(\ref{lcdasofb}), which uses the minimal number of parameters required to capture the essential features of the distribution.

\section{$\Lambda_{b}\to\Lambda_{c}$ Form Factors} \label{sec:4}
The $\Lambda_b\rightarrow \Lambda_c$ transition matrix elements induced by the vector and axial-vector currents can be decomposed into three dimensionless invariant form factors~\cite{Gutsche:2014zna, Gutsche:2015mxa}
	\begin{widetext}
		\begin{align}
			\langle  \Lambda_c(p^\prime,s^\prime)|\bar c \gamma^\mu b|\Lambda_b(p,s)\rangle=&\bar{u}_{\Lambda_c}(p^\prime,s^\prime)\left[f_1(q^2)\gamma^\mu
			-\frac{f_2(q^2)}{m_{\Lambda_b}}i\sigma^{\mu\nu}q_\nu+\frac{f_3(q^2)}{m_{\Lambda_b}}q^\mu\right]u_{\Lambda_b}(p,s),\\
			\langle  \Lambda_c(p^\prime,s^\prime)|\bar c \gamma^\mu\gamma_5 b|\Lambda_b(p,s)\rangle=&\bar{u}_{\Lambda_c}(p^\prime,s^\prime)\left[g_1(q^2)\gamma^\mu-\frac{g_2(q^2)}{m_{\Lambda_b}}i\sigma^{\mu\nu}q_\nu+\frac{g_3(q^2)}{m_{\Lambda_b}}q^\mu\right]\gamma_5u_{\Lambda_b}(p,s),
		\end{align}
	\end{widetext}
where $\sigma^{\mu\nu}=i(\gamma^\mu\gamma^\nu-\gamma^\nu\gamma^\mu)/2$, and the four-momentum transfer $q$ is constrained by the physical kinematic region $0<q^2<(m_{\Lambda_{b}}-m_{\Lambda_c})^2$.

In the previous section, we discussed the wave functions appearing in the factorization formula in Eq.~(\ref{convolution}). In this section, we calculate the hard part $H$, which includes the current operators and the necessary hard gluon exchanges connecting the current to the spectator quarks. Since the final results are expressed as integrals over the distribution variables, we present the full amplitude for each diagram, incorporating the wave functions and Sudakov factors. Given that the pQCD predictions are most reliable in the large recoil (small $q^2$) region, we evaluate the form factors at $q^2=0$ and subsequently extrapolate them over the entire kinematic range $0 \leq q^2 \leq (m_{\Lambda_b} - m_{\Lambda_c})^2$, so as to estimate the branching fractions of the semileptonic decays $\Lambda_b \to \Lambda_c \ell^- \bar{\nu}_\ell$.
	
\begin{figure*}[!htb]
		\centering
		\includegraphics[width=1.0\linewidth]{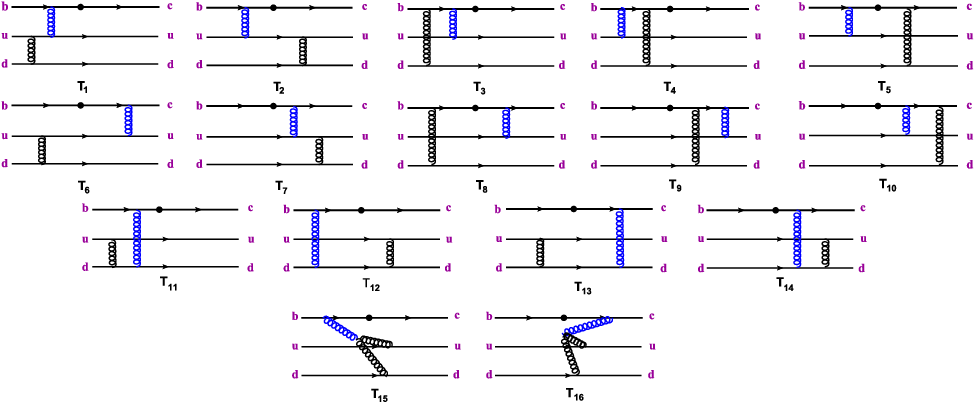}
		\caption{Feynman diagrams for the $\Lambda_{b}\xrightarrow{}\Lambda_{c}$ transition form factors with $T_i$. These diagrams are marked in sequence as $T_1,T_2$,... and $T_{16}$ in this work.}
		\label{Feynman diagram}
	\end{figure*}

In the conventional quark model, a baryon consists of three constituent quarks, making the QCD dynamics of baryon decay processes inherently more complicate than those of mesons. This complexity arises from the increased number of possible gluon exchanges, particularly within the pQCD framework, where at least two hard gluon exchanges are required at leading-order approximation. Consequently, the decay amplitudes for processes such as $\Lambda_b \to \Lambda_c$ begin at the order of $\mathcal{O}(\alpha_s^2)$ in pQCD. All factorizable Feynman diagrams contributing to this decay are illustrated in Fig.~\ref{Feynman diagram}.
	
It is noteworthy that several diagrams in Fig.~\ref{Feynman diagram} are related through the interchange of two light quarks. For instance, diagram $T_1$ can be transformed into $T_{11}$ by exchanging the up and down quarks ($u \leftrightarrow d$). As a result, the amplitude of $T_{11}$ can be obtained from that of $T_1$ by simultaneously swapping the momentum indices 2 and 3 and the spinor indices $\beta$ and $\gamma$. In contrast, diagrams ($T_{15}$ and $T_{16}$) involving the triple-gluon vertex do not contribute in this analysis, as their corresponding color factors vanish in the present case. Based on this symmetry, the following pairs of diagrams are found to be equivalent:
	\begin{eqnarray}
		T_{1}&=&T_{11},\quad T_{2}=T_{12},\quad T_{3}=T_{4},\quad T_{5}=T_{8}, \nonumber \\
		T_{6}&=&T_{13},\quad T_{7}=T_{14},\quad T_{9}=T_{10}.
	\end{eqnarray}
So, we only need to calculate the left 7 diagrams $T_1$, $T_2$, $T_3$, $T_5$, $T_6$, $T_7$ and  $T_9$.

For each diagram $T_i$, the contribution of diagram  to the form factor $F_j$ ($F=f,g$) can be evaluated at $q^2 = 0$, yielding
	\begin{eqnarray}
F_{j}^{T_{i}}\left(q^2=0\right)&=&\mathcal{C}\frac{f_{\Lambda_{b}}f_{\Lambda_{c}}\pi^3}{4 N_{c}^2}\int\left[{\cal D}x\right]\int\left[{\cal D}b\right]\int\left[{\cal D}\theta\right]\nonumber\\&&\phi_{\Lambda_{b}}\phi_{\Lambda_{c}}\alpha_{s}^{2}\left(t_{T_{i}}\right) B_{T_{i}}H^{T_{i}}_{F_j}\exp\left[-S_{T_{i}}\right],
	\end{eqnarray}
where $\mathcal{C} = 8/3$ is the color factor. The integration measure over the momentum fractions is defined as
	\begin{eqnarray}
		[\mathcal{D} x]&=&[dx][dx^{\prime}],  \\
		\left[dx\right]&=&dx_1dx_2 dx_3 \delta(1-x_1-x_2-x_3), \\
		\left[dx^{\prime}\right]&=&dx^{\prime}_1 dx^{\prime}_2 dx^{\prime}_3
		\delta(1-x^{\prime}_1-x^{\prime}_2-x^{\prime}_3),
	\end{eqnarray}
while the measure for the transverse separations reads
	\begin{eqnarray}
		[\mathcal{D} b]=b_2b_3b_2^\prime b_3^\prime db_2db_3 db_2^\prime  db_3^\prime.
	\end{eqnarray}
and the angular integration over the relative orientations in transverse ($b$-)space is given by
	\begin{eqnarray}
		[\mathcal{D} \theta]=d\theta_1 d\theta_2 d\theta_3.
	\end{eqnarray}
The hard function $B_{T_i}$ originates from the Fourier transformation of denominators of the internal propagators in diagram $T_i$, while $H^{T_i}_{F_J}$ denotes the hard scattering amplitude, which depends on the spinor structures of the valence quarks in the initial and final baryons. Explicit expressions for both $B_{T_j}$ and  $H^{T_j}$ are provided in the Appendix. Following the same procedure, we calculate all form factor components $f_i^{T_j}$ and $g_i^{T_j}$ for each contributing diagram. The complete analytic expressions are also collected in the Appendix. Finally, summing over all diagrams yields the full form factors $f_i$ and $g_i$ at $q^2 = 0$.	
	
The numerical results for the form factors are summarized in Table.~\ref{tab:result}, alongside those obtained from various theoretical approaches in the literature for comparison. The dominant theoretical uncertainties arise from the shape parameters $\beta_Q$ in the baryon LCDAs, the charm quark mass $m_c$, and the choice of the hard scale $t$. In our analysis, we estimate these uncertainties by varying $\beta_Q$ and $m_c$ within a $10\%$ range, and by varying the hard scale in the range $0.8t$ to $1.2t$. Among these sources, the most significant uncertainty originates from the baryon LCDAs, contributing up to $23\%$  in magnitude. This sizable uncertainty highlights the necessity of better constraining the non-perturbative parameters in the baryon distribution amplitudes to improve the precision of theoretical predictions.
	
	\begin{table*}[!htb]
		\centering
		\caption{Results of the form factor at $q^2=0$ of this work are compared with those of other methods.}
		\label{tab:result}
		\renewcommand{\arraystretch}{1.2}
		\begin{tabular}{ccccccc}
			\hline\hline
			&$f_{1}\left(0\right)$  & $f_{2}\left(0\right)$ &$f_{3}\left(0\right)$ &$g_{1}\left(0\right)$ &$g_{2}\left(0\right)$ &$g_{3}\left(0\right)$   \\ \hline
			This work         & $0.499^{+0.091}_{-0.110}$              & $0.083^{+0.026}_{-0.027}$              & $-0.086^{+0.029}_{-0.022}$             & $0.504^{+0.076}_{-0.119}$            & $0.091^{+0.027}_{-0.030}$             & $-0.088^{+0.030}_{-0.020}$        \\
			LCSR\cite{Miao:2022bga}     &$0.583$     &$0.144$    &$-0.115$    & $0.583$   &$0.114$       &$-0.115$     \\
			QCDSR\cite{Zhao:2020mod}   & $0.467$              & $0.123$              & $0.022$              & $0.434$            & $-0.036$            & $-0.160$        \\
			LFQM\cite{Zhu:2018jet}          & $0.500$              & $0.098$              & $-0.009$             & $0.509$            & $0.015$             & $-0.085$        \\
			LFQM\cite{Li:2021qod} & $0.50\pm{0.05}$  & $0.12\pm{0.01}$  & $-0.04\pm{0.00}$ & $0.49\pm{0.05}$ & $0.02\pm{0.00}$ & $-0.15\pm{0.02}$ \\
			RQM\cite{Faustov:2016pal}           & $0.526$              & $0.136$              & $0.075$              & $0.504$            & $-0.026$            & $-0.251$        \\
			CCQM\cite{Gutsche:2015mxa}          & $0.549$              & $0.110$              & $-0.023$             & $0.542$            & $0.018$             & $-0.123$        \\
			LQCD\cite{Detmold:2015aaa}          & $0.418\pm0.161$              & $0.099$              & $-0.075$             & $0.378\pm0.102$            & $0.004$             & $-0.205$        \\
			LFQM\cite{Ke:2007tg}          & $0.506$              & $0.099$              & $-$                  & $0.501$            & $0.008$             & $-$             \\
			LFQM\cite{Zhao:2018zcb}          & $0.670$              & $0.132$              & $-$                  & $0.656$            & $0.012$             & $-$             \\
			LFQM\cite{Chua:2019yqh} & $0.474^{+0.069}_{-0.072}$ & $0.108^{+0.020}_{-0.019}$ & $0.049^{+0.014}_{-0.015}$& $0.468^{+0.067}_{-0.070}$ & $-0.050^{+0.011}_{-0.008}$ & $-0.118^{+0.017}_{-0.015}$ \\
			LFQM\cite{Ke:2019smy}          & $0.488$              & $0.180$              & $-$                  & $0.470$            & $0.048$             & $-$             \\
			pQCD\cite{Zhang:2022iun}   & $0.440^{+0.136}_{-0.173}$ & $-$ &$-$  & $0.443^{+0.143}_{-0.181}$  & $-$ & $-$ \\
			\hline\hline
		\end{tabular}
	\end{table*}
	
In order to calculate the observables of semileptonic decays $\Lambda_{b}\to\Lambda_{c}\ell^-\bar{\nu}_{\ell}$ ($\ell=e,\mu,\tau$), the behaviors of the form factors in the whole kinematic range $0\leq q^2\leq \left(m_{\Lambda_b}-m_{\Lambda_c}\right)^2$ are required. Now, we shall extrapolate from the low $q^2$ to the higher $q^2$ region for all form factors. In the literature, there are lots of parameterizations, such as single-pole model and double pole model. In this work, we shall adopt the simplified $z$-series parameterization \cite{Miao:2022bga}
	\begin{eqnarray}
		f_{i}\left(q^2\right)=\frac{f_{i}\left(0\right)}{1-q^2/m^2_{B_{c}^{*}\left(1^-\right)}}
		\Big\{1+a_{1}\left[z\left(q^2,0\right)-z\left(0,t_{0}\right)\right]\Big\},\nonumber\\
		g_{i}\left(q^2\right)=\frac{g_{i}\left(0\right)}{1-q^2/m^2_{B_{c}^{*}\left(1^+\right)}}
		\Big\{1+b_{1}\left[z\left(q^2,0\right)-z\left(0,t_{0}\right)\right]\Big\},
	\end{eqnarray}
where the parameter $z$ is defined as
	\begin{eqnarray}
		z\left(q^2,t_0\right)=\frac{\sqrt{t_+-q^2}-\sqrt{t_+-t_0}}{\sqrt{t_+-q^2}+\sqrt{t_+-t_0}}
	\end{eqnarray}
with $t_{\pm}=\left(m_{\Lambda_b}\pm m_{\Lambda_c}\right)^2$ and $t_0=t_+\left(1-\sqrt{1-t_-/t_+}\right)$. The masses of $B_{c}^{*}\left(1^-\right)$ and $B_{c}^{*}\left(1^+\right)$ appear in the pole factor but are not measured. There are several theoretical estimations of these masses, and we adopt $m_{B^{*}\left(1^-\right)}=6.336 ~\rm{GeV}$  and $m_{B^{*}\left(1^+\right)}=6.745~\rm{GeV}$ \cite{Bernlochner:2018bfn}.
	
	\begin{figure*}[!htb]
	\centering
	\includegraphics[width=0.45\linewidth]{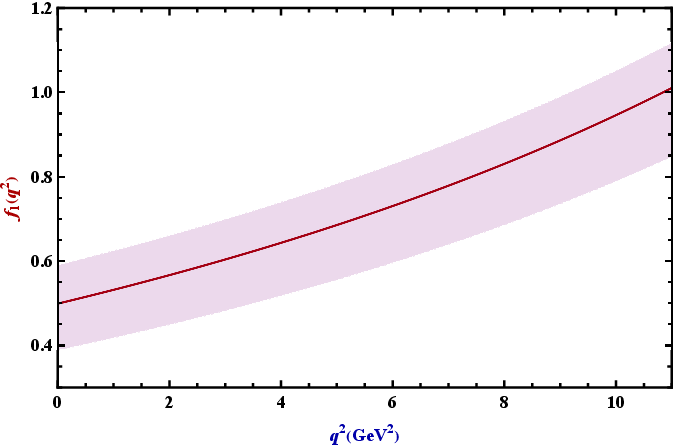}
	\includegraphics[width=0.45\linewidth]{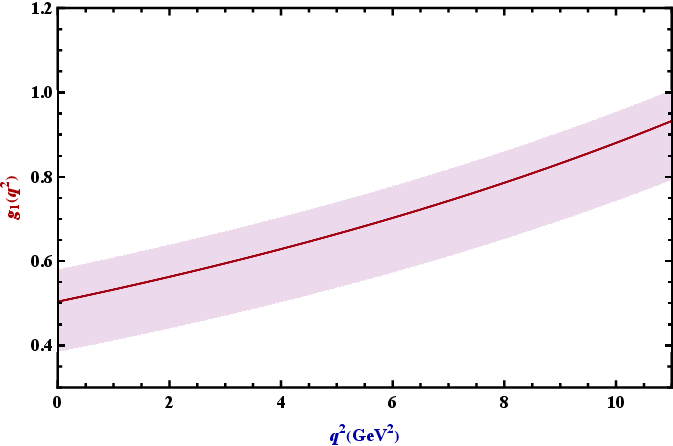}
	\includegraphics[width=0.45\linewidth]{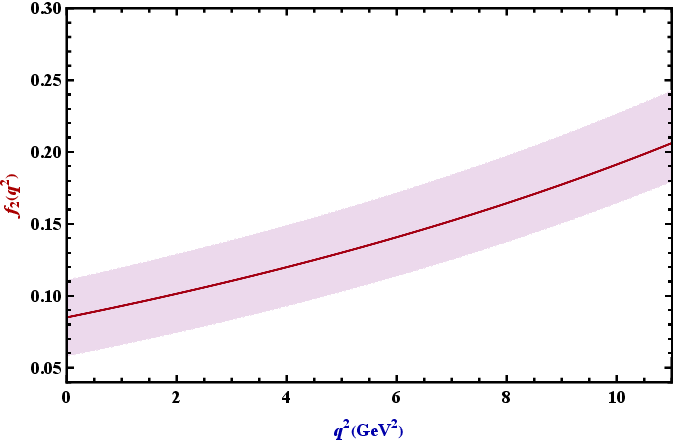}
	\includegraphics[width=0.45\linewidth]{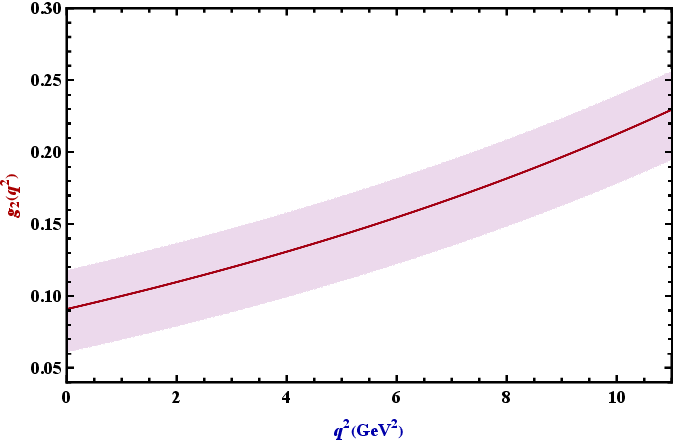}
	\includegraphics[width=0.45\linewidth]{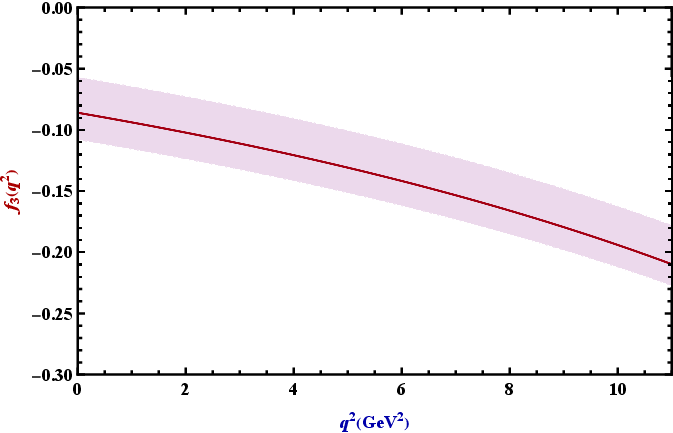}
	\includegraphics[width=0.45\linewidth]{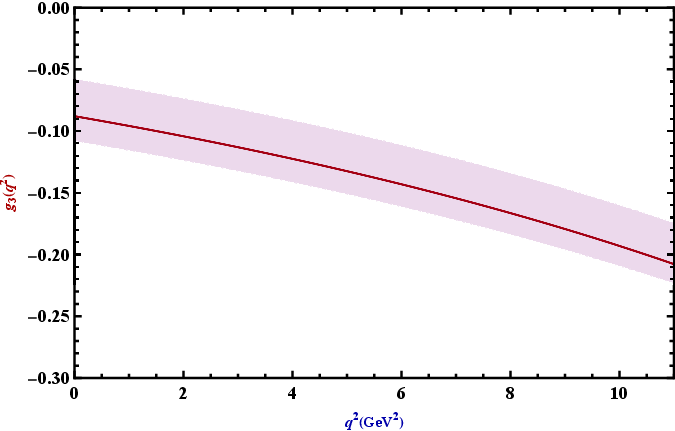}
	
	\caption{The $\Lambda_{b}\to\Lambda_{c}$ form factors induced by the (axial)-vector currents fitted to the z-series parameterizations.}
	\label{form factors}
\end{figure*}
	
	\begin{table*}[!htb]
		\centering
		\caption{Fitted result of $a_1,b_1$ for the form factors.}
		\label{tab:fit}
		\renewcommand{\arraystretch}{1.4}
		\setlength{\abovecaptionskip}{1cm}
		\setlength{\belowcaptionskip}{-0.5cm}
		\setlength{\tabcolsep}{40pt}
		\begin{tabular}{cccc}
			\hline\hline
			$F$             &$F\left(0\right)$  & $a_1$     &$b_1$  \\ \hline
			$f_{1}$         & $0.499^{+0.091}_{-0.110}$              & $-9.73^{+1.98}_{-2.32}$              & $ $                   \\
			$f_{2}$     &$0.083^{+0.026}_{-0.027}$     &$-17.37^{+3.75}_{-5.21}$    &$ $                                          \\
			$f_{3}$     &$-0.086^{+0.029}_{-0.022}$              & $-15.96^{+5.04}_{-5.49}$              & $ $                    \\
			$g_{1}$       & $0.504^{+0.076}_{-0.119}$              & $ $              & $-8.34^{+1.83}_{-3.28}$                 \\
			$g_{2}$     & $0.091^{+0.027}_{-0.030}$  & $ $  & $-18.91^{+5.49}_{-6.09}$               \\
			$g_{3}$        & $-0.088^{+0.030}_{-0.020}$              & $ $              & $-16.35^{+4.61}_{-5.37}$                 \\
			\hline\hline
		\end{tabular}
	\end{table*}
	
It is known to us that the form factors based on pQCD are reliable only in the low $q^2$ region, and in the large $q^2$ region, the applicability of the pQCD framework may break down. In additional, we also know that in the large $q^2$ region, the results based on the lattice QCD are reliable. In ref.\cite{Detmold:2015aaa}, the lattice QCD results for the 6 form factors are published, from which we obtain the numerical results at $q_{max}^2=(m_{\Lambda_b}- m_{\Lambda_c})^2$ as
	\begin{eqnarray}
		f_1(q_{max}^2)=1.011^{+0.106}_{-0.163},&&
		g_1(q_{max}^2)=0.932^{+0.074}_{-0.139},\nonumber\\
		f_2(q_{max}^2)=0.206^{+0.037}_{-0.026},&&
		g_2(q_{max}^2)=0.230^{+0.026}_{-0.035},\nonumber\\
		f_3(q_{max}^2)=-0.204^{+0.026}_{-0.023},&&
		g_3(q_{max}^2)=-0.207^{+0.044}_{-0.016}. \label{LQCD}
	\end{eqnarray}
Thus, in the low $q^2$ region from 0 to $m_{\tau}^2$, $m_{\tau}$ being the $\tau$ lepton mass, we adopt results of pQCD approach, while in the large $q^2$ region, we use the results from Lattice QCD, as shown in Eq.(\ref{LQCD}). Within the data, the fitted results of $a_1,b_1$ are presented in Table \ref{tab:fit}. The $q^2$ behaviors of the form factors in the allowed region are also illustrated in Fig.\ref{form factors}. As it is expected from weak decays, the form factors demonstrate a good behavior that their magnitudes grow gradually with increasing $q^2$. $f_{1}(q^2)$ and $g_1(q^2)$ show similar $q^2$ dependence and dominate over other four form factors.
	
\section{Analysis of $\Lambda_{b}\to\Lambda_{c} \ell^-\bar \nu_\ell $} \label{sec:5}
In the section, we shall explore the phenomenological applications of the obtained $\Lambda_b\to\Lambda_c$ form factors, which serve as essential inputs for the theoretical description of the semileptonic decay $\Lambda_b \to \Lambda_c \ell^- \bar{\nu}_\ell$. Within the SM, this decay is typically viewed as proceeding through an intermediate off-shell $W^{*-}$ boson: $\Lambda_b \to \Lambda_c W^{*-}$, followed by the subsequent decay $W^{*-} \to \ell^- \bar{\nu}_\ell$. The off-shell $W^{*-}$ boson can carry four helicity states: $\lambda_W = \pm 1$, $0$ (with total spin $J=1$), and a scalar polarization corresponding to $J=0$, also labeled by $\lambda_W=0$. To distinguish between these two states with $\lambda_W=0$, we adopt the notation $\lambda_W=0$ for the longitudinal polarization ($J=1$) and $\lambda_W=t$ for the scalar (time-like) polarization ($J=0$).
	
In the rest frame of the $\Lambda_b$ baryon, with the $z$-axis defined along the momentum direction of the virtual $W^{*-}$ boson, the polarization vectors of the $W^{*-}$ are given by:
	\begin{eqnarray}
		\varepsilon^\mu(\pm)&=&\frac{1}{\sqrt{2}}(0,1,\mp i,0);\nonumber\\
		\varepsilon^\mu(0)&=&-\frac{1}{\sqrt{q^2}}(q_z,0,0,q_0); \nonumber\\
		\varepsilon^\mu(t)&=&-\frac{q^\mu}{\sqrt{q^2}};
	\end{eqnarray}
where $q^\mu$ denotes the four-momentum of the off-shell $W^{*-}$ boson. These polarization vectors satisfy the following orthonormality and completeness relations:
	\begin{eqnarray}
		\varepsilon^{*\mu}(m)\epsilon_\mu(n)=g_{mn},\,\,
		\sum_{m,n}\varepsilon^{*\mu}(m)\varepsilon^\nu(n)g_{mn}=g^{\mu\nu}
	\end{eqnarray}
with $g_{mn}={\rm diag}(+,-,-,-)$ for $m,n=t,\pm,0$. Because the current is not conserved in SM, it consists of a superposition of a spin-1 and a spin-0 component where the $J^{P}$ content of the vector current $J^V_{\mu}$ and the axial vector current $J^A_{\mu}$ are $(0^{+},1^{-})$ and $(0^{-},1^{+})$, respectively.
	
In the calculation of the $\Lambda_b \to \Lambda_c \ell^- \bar{\nu}_\ell$ decay, the total transition matrix element can be factorized into a leptonic part and a hadronic part, each of which is not Lorentz invariant by itself. However, upon inserting the completeness relations of the intermediate off-shell $W^{*-}$ boson, both the hadronic and leptonic contributions become Lorentz invariant. This allows us to choose the most convenient reference frames for evaluating each part of the amplitude. Specifically, the hadronic transition is analyzed in the rest frame of the initial $\Lambda_b$ baryon, while the leptonic part is evaluated in the rest frame of the virtual $W^{*-}$ boson. Based on this setup, we proceed to calculate the helicity amplitudes for the $\Lambda_b \to \Lambda_c W^{*-}$ transition as follows:
	\begin{eqnarray}
		H^{V(A)}_{\lambda_{\Lambda_{c}},\lambda_{W}}&=&\varepsilon^{*\mu}\left(\lambda_{W^{-}}\right)\left<\Lambda_{c}(\lambda_{\Lambda_{c}})|V\left(A\right)|\Lambda_{b}(\lambda_{\Lambda_{b}})\right>,
	\end{eqnarray}
where $\lambda_{\Lambda_{b}}, \lambda_{\Lambda_{c}}, \lambda_{W}$ denote the helicity of the $\Lambda_{b}$ baryon, $\Lambda_{c}$ baryon, and off-shell $W^{*-}$, respectively. Due to the conservation of angular momentum,  $\lambda_{1}=-\lambda_{2}+\lambda_{W}$ is satisfied. The various helicity amplitudes are then expressed, in terms of the $\Lambda_{b}\xrightarrow{}\Lambda_{c}$ form factors, as \cite{Azizi:2019tcn,Azizi:2018axf,Gutsche:2015mxa,Shivashankara:2015cta,Dutta:2015ueb,Zwicky:2013eda,Hiller:2021zth}
	\begin{widetext}
		\begin{align}
			H^{V}_{\frac{1}{2},0}&=\frac{\sqrt{\left(m_{\Lambda_{b}}-m_{\Lambda_{c}}\right)^{2}-q^2}}{\sqrt{q^2}}\left[\left(m_{\Lambda_{b}}+m_{\Lambda_{c}}\right)f_{1}\left(q^2\right)+\frac{q^2}{m_{\Lambda_{b}}}f_{2}\left(q^2\right)\right],\\
			H^{A}_{\frac{1}{2},0}&=\frac{\sqrt{\left(m_{\Lambda_{b}}+m_{\Lambda_{c}}\right)^{2}-q^2}}{\sqrt{q^2}}\left[\left(m_{\Lambda_{b}}-m_{\Lambda_{c}}\right)g_{1}\left(q^2\right)-\frac{q^2}{m_{\Lambda_{b}}}g_{2}\left(q^2\right)\right],\\
			H^{V}_{\frac{1}{2},1}&=\sqrt{2\left[\left(m_{\Lambda_{b}}-m_{\Lambda_{c}}\right)^{2}-q^2\right]}\left[-f_{1}\left(q^2\right)-\frac{m_{\Lambda_{b}}+m_{\Lambda_{c}}}{m_{\Lambda_{b}}}f_{2}\left(q^2\right)\right],\\
			H^{A}_{\frac{1}{2},1}&=\sqrt{2\left[\left(m_{\Lambda_{b}}+m_{\Lambda_{c}}\right)^{2}-q^2\right]}\left[-g_{1}\left(q^2\right)+\frac{m_{\Lambda_{b}}-m_{\Lambda_{c}}}{m_{\Lambda_{b}}}g_{2}\left(q^2\right)\right],\\
			H^{V}_{\frac{1}{2},t}&=\frac{\sqrt{\left(m_{\Lambda_{b}}+m_{\Lambda_{c}}\right)^{2}-q^2}}{\sqrt{q^2}}\left[\left(m_{\Lambda_{b}}-m_{\Lambda_{c}}\right)f_{1}\left(q^2\right)+\frac{q^2}{m_{\Lambda_{b}}}f_{3}\left(q^2\right)\right],\\
			H^{A}_{\frac{1}{2},t}&=\frac{\sqrt{\left(m_{\Lambda_{b}}-m_{\Lambda_{c}}\right)^{2}-q^2}}{\sqrt{q^2}}\left[\left(m_{\Lambda_{b}}+m_{\Lambda_{c}}\right)g_{1}\left(q^2\right)-\frac{q^2}{m_{\Lambda_{b}}}g_{3}\left(q^2\right)\right],
		\end{align}
	\end{widetext}
where the subscript $t$ refers to the temporal component of the helicities of the off-shell $W^{-}$ boson \cite{Gutsche:2015mxa}. The amplitudes with negative helicities can be obtained by
	\begin{eqnarray}
		H^{V}_{-\lambda_{\Lambda_{c}},-\lambda_{W}}&=&H^{V}_{\lambda_{\Lambda_{c}},\lambda_{W}}, \nonumber\\ H^{A}_{-\lambda_{\Lambda_{c}},-\lambda_{W}}&=&-H^{A}_{\lambda_{\Lambda_{c}},\lambda_{W}}.
	\end{eqnarray}
Then, we have the helicity amplitude as
	\begin{eqnarray}
		H_{\lambda_{\Lambda_{c}},\lambda_{{W}}}=H^{V}_{\lambda_{\Lambda_{c}},\lambda_{W}}-H^{A}_{\lambda_{\Lambda_{c}},\lambda_{W}},
	\end{eqnarray}
Since the leptonic sector is free from hadronic QCD complications, the corresponding amplitudes can be evaluated analytically. Detailed expressions are available in Ref.~\cite{Ivanov:2016qtw}.
	
The differential angular distributions for the semileptonic decays $\Lambda_{b}\to\Lambda_{c}\ell^-\bar{\nu}_{\ell}$ are given by \cite{Li:2021qod,Azizi:2019tcn,Miao:2022bga}
	\begin{widetext}
		\begin{align}\label{eq:dw}
			\frac{d\Gamma\left(\Lambda_{b}\to\Lambda_{c}\ell^{-}\bar{\nu}_{\ell}\right)}{dq^{2}d\cos{\theta_{\ell}}}
			=\frac{G^{2}_{F}\left|V_{cb}\right|^{2}q^{2}\left|\Vec{p}_{c}\right|}{512\pi^{3}m^{2}_{\Lambda_{b}}}\left(1-\frac{m^{2}_{\ell}}{q^2}\right)^{2} \left(A_{1}+\frac{m^{2}_{\ell}}{q^2}A_{2}\right),
		\end{align}
	\end{widetext}
where $m_{\ell}$ is the lepton mass $(\ell=e,\mu,\tau)$, $G_{F}$ is the Fermi constant, $V_{cb}$ is the CKM matrix element, $\theta_{\ell}$ stands for the angle between the charged lepton and the final $\Lambda_{c}$ baryon in the $\Lambda_{b}$ baryon rest frame and $\vec{p}_{c}$ is the three-momentum of the $\Lambda_{c}$ baryon. The amplitudes $A_{1},A_{2}$ are defined as
	\begin{eqnarray}
		A_{1}&=&2\sin^{2}\theta_{\ell}\left(H^{2}_{\frac{1}{2},0}+H^{2}_{-\frac{1}{2},0}\right)+\left(1-\cos{\theta_{\ell}}\right)^{2}H^{2}_{\frac{1}{2},1}\nonumber\\&&+\left(1+\cos{\theta_{\ell}}\right)^{2}H^{2}_{-\frac{1}{2},-1},\\
		A_{2}&=&2\cos^{2}\theta_{\ell}\left(H^{2}_{\frac{1}{2},0}+H^{2}_{-\frac{1}{2},0}\right)+2\left(H^{2}_{\frac{1}{2},t}+H^{2}_{-\frac{1}{2},t}\right) \nonumber\\
		&&+\sin^{2}\theta_{\ell}\left(H^{2}_{\frac{1}{2},1}+H^{2}_{-\frac{1}{2},1}+H^{2}_{-\frac{1}{2},-1}\right)\nonumber\\
		&&+\-4\cos{\theta_{\ell}}\left(H_{\frac{1}{2},t}H_{\frac{1}{2},0}+H_{-\frac{1}{2},t}H_{-\frac{1}{2},0}\right).
	\end{eqnarray}
	
Integration of Eq.~(\ref{eq:dw}) over $\cos\theta_\ell$ yields the $q^2$-dependent differential decay width:
	\begin{eqnarray}
		\label{eq:ddw}
		\frac{d\Gamma\left(\Lambda_{b}\to\Lambda_{c}\ell^{-}\Bar{v}_{\ell}\right)}{dq^2}=\int^{1}_{-1}\frac{d\Gamma\left(\Lambda_{b}\xrightarrow{}\Lambda_{c}\ell^{-}\Bar{v}_{\ell}\right)}{dq^{2}d\cos{\theta_{\ell}}}d\cos{\theta_{\ell}}.
	\end{eqnarray}
Since the masses of the electron and muon are negligible compared to those of the initial and final baryons, the SM predictions for the $e$ and $\mu$ channels are nearly identical. Therefore, only the results for the $\mu$ and $\tau$ modes are discussed in the following.
	\begin{table*}[!htb]
		\centering
		\caption{Theoretical predictions for the $\Lambda_{b}\xrightarrow{}\Lambda_{c}$ semileptonic decay branching fractions and the LFU ratio.}
		\label{tab:br}
		\renewcommand{\arraystretch}{1.4}
		\setlength{\abovecaptionskip}{1cm}
		\setlength{\belowcaptionskip}{-0.5cm}
		\setlength{\tabcolsep}{25pt}
		\begin{tabular}{cccc}
			\hline \hline
			Model& $\mathcal{B}\left(\Lambda_{b}\xrightarrow{}\Lambda_{c}\mu\bar{\nu}_{\mu}\right)$ & $\mathcal{B}\left(\Lambda_{b}\xrightarrow{}\Lambda_{c}\tau\bar{\nu}_{\tau}\right)$ & $\mathcal{R}_{\Lambda_{c}}$ \\ \hline
			This work & $\left(5.8^{+1.5}_{-2.0}\right)\times10^{-2}$ & $\left(1.7^{+0.4}_{-0.5}\right)\times10^{-2}$ & $0.29^{+0.12}_{-0.11}$ \\
			RQM\cite{Faustov:2016pal} & $6.48\times10^{-2}$ & $2.03\times10^{-2}$ & 0.313 \\
			CCQM\cite{Zhu:2018jet} & $6.9\times10^{-2}$ & $2.0\times10^{-2}$ & 0.29 \\
			QCDSR\cite{Gutsche:2015mxa} & $5.57\times10^{-2}$ & $1.54\times10^{-2}$ & 0.28 \\
			LQCD\cite{Detmold:2015aaa} & ... &... & 0.3318 \\
			HQET@NLP\cite{Bernlochner:2018bfn} & ... &... & 0.324$\pm$0.004 \\
			\hline \hline
		\end{tabular}
	\end{table*}
	
To probe potential effects beyond the SM, it is essential to evaluate LFU, which is defined as Eq.~(\ref{RLambdac}).
Our predictions for the branching fractions and LFU  parameters are presented in Table~\ref{tab:br}, along with results from other theoretical approaches for comparison. Due to the heavier mass of the $\tau$ lepton, the corresponding decay channel is subject to kinematic suppression, leading to an expected LFU ratio $R_{\Lambda_c}$ smaller than one. From the table, it is found that for $\mathcal{R}_{\Lambda_{c}}$ our predictions is lower than the estimations of Refs.~\cite{Detmold:2015aaa,Faustov:2016pal}, but slightly higher than value reported in Ref.~\cite{Gutsche:2015mxa}. The LHCb collaboration has recently reported a measurement of the LFU ratio in the $\Lambda_{b}\to \Lambda_{c}\ell^-\bar{\nu}_{\ell}$ decay channel, obtaining $\mathcal{R}_{\Lambda{c}}=0.242\pm0.026 \text{(stat)} \pm0.040 \text{(syst)}\pm0.059 \text{(ext)}$ \cite{LHCb:2022piu}, where the uncertainties correspond to statistical, systematic, and external branching fraction contributions, respectively. It is found that our result is slightly larger than the experimental data, corresponding to a tension of approximately $0.35\sigma$. And we also note that this experimental result exhibits a $1.3\sigma$ tension with the SM prediction of $\mathcal{R}_{\Lambda_c}=0.324 \pm 0.004$ \cite{Bernlochner:2018bfn}, which incorporates both lattice QCD calculations and measured spectral information. This observed discrepancy may suggest a potential violation of LFU in semileptonic $\Lambda_b \to \Lambda_c \ell^- \bar{\nu}_{\ell}$ decays. Our theoretical predictions provide complementary insights into the study of LFU in the $b \to c \ell \bar{\nu}_{\ell}$ transitions and may serve as a useful reference for future experimental investigations.
	
It is worth emphasizing that the predicted branching fractions are subject to sizable theoretical uncertainties. These mainly arise from the modeling of hadronic form factors, uncertainties in the parameters of the light-cone distribution amplitudes (LCDAs), and the omission of higher-order QCD and power corrections. Reducing these uncertainties—through improved nonperturbative QCD inputs, refined nonperturbative parameter extraction, and the inclusion of radiative and subleading contributions—will be essential for enhancing the precision and reliability of theoretical predictions in future studies.
	
	\begin{table*}[!htb]
		\centering
		\caption{Theoretical predictions for the $\Lambda_{b}$ semileptonic decay parameters with available theoretical approaches.}
		\label{tab:observables}
		\renewcommand{\arraystretch}{1.4}
		\setlength{\abovecaptionskip}{1cm}
		\setlength{\belowcaptionskip}{-0.5cm}
		\setlength{\tabcolsep}{15pt}
		\begin{tabular}{ccccccc}
			\hline \hline
			Model& $\ell$ & $\left<\Gamma\right>$ & $\left<A_{FB}\right>$ &$\left<P_{B}\right>$  & $\left<P_{\ell}\right>$ & $\left<C_{F}\right>$  \\ \hline
			& $e$ & $26$ & $0.193^{+0.006}_{-0.001}$ & $-0.928^{+0.001}_{-0.014}$ & $-1.0$ & $-0.751^{+0.002}_{-0.007}$ \\
			This work & $\mu$ & $26$ & $0.182^{+0.009}_{-0.003}$ & $-0.930^{+0.001}_{-0.011}$ & $-1.0$  & $-0.727^{+0.002}_{-0.007}$ \\
			& $\tau$ & $7.7$ & $-0.076^{+0.002}_{-0.001}$ & $-0.630^{+0.012}_{-0.002}$ & $-0.173^{+0.007}_{-0.001}$ & $-0.088^{+0.001}_{-0.001}$ \\
			& $e$ & $29.4$ & $0.195$ & $-0.80$ &... & $-0.57$ \\
			RQM\cite{Faustov:2016pal} & $\mu$ & $29.0$ & $0.189$ & $-0.80$ &... & $-0.55$\\
			&  $\tau$ & $9.1$ & $-0.021$ & $-0.71$ & $...$ & $-0.09$  \\
			& $e$ & $32.0$ & $0.36$ & $-0.82$ & $-1.00$ & $-0.63$ \\
			CCQM\cite{Gutsche:2015mxa}&$\mu$ & $...$ & ... &... &... &...\\
			& $\tau$ & $9.40$ & $-0.077$ & $-0.72$ & $-0.32$ & $-0.10$ \\
			& $e$ & $...$ & $0.18$ & $-0.81$ & $-1.00$ &...\\
			LFQM\cite{Li:2021qod} & $\mu$ & $...$ & $0.17$ & $-0.81$ & $-0.98$ & $...$\\
			& $\tau$ &... & $-0.08$ & $-0.77$ & $-0.24$ &...\\
			\hline \hline
		\end{tabular}
	\end{table*}
	
In particular, the forward-backward asymmetry ($A_{FB}$) of the leptonic sector is of significant interest, as it exhibits reduced sensitivity to hadronic form factor uncertainties. It is defined as follows:
	\begin{eqnarray}
		\label{eq:AFB}
		A_{FB}^\ell\left(q^2\right)=\frac{\int_{0}^{1}\frac{d\Gamma}{dq^{2}d\cos{\theta_{\ell}}}d\cos{\theta_{\ell}}-\int_{-1}^{0}\frac{d\Gamma}{dq^{2}d\cos{\theta_{\ell}}}d\cos{\theta_{\ell}}}{\int_{0}^{1}\frac{d\Gamma}{dq^{2}d\cos{\theta_{\ell}}}d\cos{\theta_{\ell}}+\int_{-1}^{0}\frac{d\Gamma}{dq^{2}d\cos{\theta_{\ell}}}d\cos{\theta_{\ell}}},
	\end{eqnarray}
In addition, other observables, such as the final state hadron polarization $(P_{B})$ and lepton polarization $(P_{\ell})$, are defined as
	\begin{eqnarray}
		P_{\ell}\left(q^2\right)&=&\frac{d\Gamma^{\lambda_{\ell}=1/2}/dq^2-d\Gamma^{\lambda_{\ell}=-1/2}/dq^2}{d\Gamma^{\lambda_{\ell}=1/2}/dq^2+d\Gamma^{\lambda_{\ell}=-1/2}/dq^2},	\label{eq:PL}\\
		P_{B}\left(q^2\right)&=&\frac{d\Gamma^{\lambda_{\Lambda_{c}}=1/2}/dq^2-d\Gamma^{\lambda_{\Lambda_{c}}=-1/2}/dq^2}{d\Gamma^{\lambda_{\Lambda_{c}}=1/2}/dq^2+d\Gamma^{\lambda_{\Lambda_{c}}=-1/2}/dq^2},  	\label{eq:PB}
	\end{eqnarray}
and the differential widths with definite polarization of the final state can be written as
	\begin{widetext}
		\begin{align}
			\frac{d\Gamma^{\lambda_{\Lambda_{c}}=1/2}}{dq^2}=&\frac{4m_{\ell}^2}{3q^2}\left(H^{2}_{1/2,1}+H^{2}_{1/2,0}+3H^{2}_{1/2,t}\right)+\frac{8}{3}\left(H^{2}_{1/2,0}+H^{2}_{1/2,1}\right), \\
			\frac{d\Gamma^{\lambda_{\Lambda_{c}}=-1/2}}{dq^2}=&\frac{4m_{\ell}^2}{3q^2}\left(H^{2}_{-1/2,-1}+H^{2}_{-1/2,0}+3H^{2}_{-1/2,t}\right)+\frac{8}{3}\left(H^{2}_{-1/2,0}+H^{2}_{-1/2,-1}\right), \\
			\frac{d\Gamma^{\lambda_{\ell}=1/2}}{dq^2}=&\frac{m_{\ell}^{2}}{q^2}\left[\frac{4}{3}\left(H^{2}_{1/2,1}+H^{2}_{1/2,0}+H^{2}_{-1/2,-1}+H^{2}_{-1/2,0}\right)+4\left(H^{2}_{1/2,t}+H^{2}_{-1/2,t}\right)\right],    \\
			\frac{d\Gamma^{\lambda_{\ell}=-1/2}}{dq^2}=&\frac{8}{3}\left(H^{2}_{1/2,1}+H^{2}_{1/2,0}+H^{2}_{-1/2,-1}+H^{2}_{-1/2,0}\right).
		\end{align}
	\end{widetext}
	
We also find that the differential angular distribution contains a large number of $\cos{\theta_{\ell}}$ terms. In order to better describe $\cos{\theta_{\ell}}$, The convexity parameter $C_{F}$ is defined as the second derivative of the normalized angular distribution
	\begin{eqnarray}\label{eq:CF}
		C_{F}^{\ell}\left(q^2\right)=\frac{1}{d\Gamma/dq^2}\frac{d^2}{d\cos^{2}{\theta}}\left(\frac{d^{2}\Gamma}{dq^{2}d\cos{\theta}}\right).
	\end{eqnarray}
In fact, it helps to disentangle the contributions from different helicity amplitudes. Since $C_{F}$can be extracted from the shape of the angular distribution, it's a robust observable for comparing experimental data with theoretical predictions. Deviations from SM expectations in $C_{F}^\ell$ may indicate new physics in the semileptonic decay processes.
	
In Table.~\ref{tab:observables}, we present our predictions for the averaged values of several key observables: the total decay width $\langle \Gamma\rangle$ (in units of $10^{-15}$ GeV), the forward-backward asymmetry of the charged lepton $\langle A_{FB}\rangle$, the polarization of the final-state baryon $\langle P_{B}\rangle$, the polarization of the charged lepton $\langle P_{\ell}\rangle$, and the lepton-side convexity parameter $\langle C_{F}^\ell\rangle$. These observables exhibit reduced sensitivity to model-dependent uncertainties, as many theoretical errors cancel in the ratios of helicity amplitudes.  Significant variations are observed when transitioning from the light lepton ($e/\mu$) to the $\tau$ channel, primarily due to the sizable mass of the $\tau$ lepton. Notably, the forward-backward asymmetry $\langle A_{FB}\rangle$ undergoes a sign reversal in the $\tau$ mode, reflecting the pronounced influence of helicity suppression effects. We further compare our results for the integrated observables with those obtained from other theoretical frameworks~\cite{Li:2021qod, Faustov:2016pal, Gutsche:2015mxa}, as also summarized in Table.~\ref{tab:observables}. A good level of consistency is found among all theoretical predictions, which provides a promising basis for future experimental validation.

	\begin{figure}[!htb]
	\centering
	\includegraphics[width=0.8\linewidth]{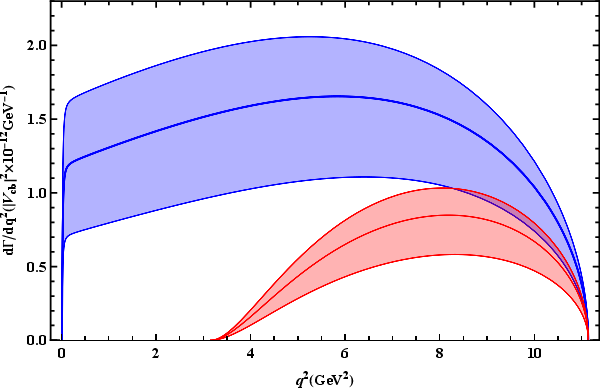}
	\caption{The $q^2$-dependence of the differential widths for the $\Lambda_{b}\to\Lambda_{c}\ell\bar{\nu}_{\ell}$ decays with $\ell=\mu,\tau$. The blue (solid) and pink (solid) regions refer to the  $\Lambda_{b}\to\Lambda_{c}\mu^-\bar{\nu}_{\mu}$ and $\Lambda_{b}\to\Lambda_{c}\tau^-\bar{\nu}_{\tau}$, respectively, and the uncertainty bands are obtained byadding all separate theory uncertainties in quadrature.}
	\label{fig:dw}
\end{figure}
	
Using the analytical expressions of the observables derived in the previous sections, we plot their dependence on the momentum transfer squared $q^2$, as illustrated in the figure below. Figure.~\ref{fig:dw} shows the $q^2$-dependence of the differential decay widths for the processes  $\Lambda_{b}\to\Lambda_{c}\ell^-\bar{\nu}_{\ell}$, with $\ell=\mu,\tau$. According to Eq.~(\ref{eq:ddw}), in the large-recoil region (i.e.,  $q^2\to m_{\ell}^2$), the differential width displays a step-like behavior. Conversely, in the zero-recoil limit, the differential width vanishes.

\begin{figure}[!htb]
		\centering
		\includegraphics[width=0.8\linewidth]{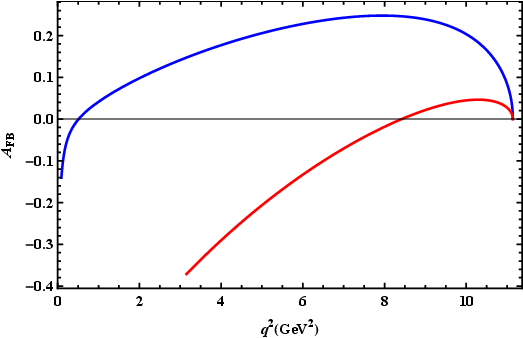}
		\caption{The $q^2$-dependence of the lepton-side forward-backward asymmetry $A_{FB}^\ell(q^2)$ for the $\mu^{-}$(blue) and $\tau^{-}$(red) modes.}
		\label{fig:AFB}
	\end{figure}

Figure.~\ref{fig:AFB} shows the $q^2$-dependence of the lepton-side forward-backward asymmetry $A_{FB}^\ell(q^2)$, as defined in Eq.~(\ref{eq:AFB}). At zero recoil,  $A^\ell_{FB}$ vanishes for both the $\mu$ and $\tau$ channels due to the helicity relation $H^{2}_{1/2,1}=H^{2}_{-1/2,-1}$. In the large-recoil limit, $A_{FB}^\mu(q^2)$ also tends to zero, reflecting the dominance of longitudinal contributions to the decay rate. Notably, the behavior of $A_{FB}(q^2)$ differs significantly between the two lepton flavors: while $A_{FB}^\mu(q^2)$ almost remains positive across the entire $q^2$ spectrum, $A_{FB}^\tau(q^2)$ becomes negative shortly after zero recoil (i.e.,  $q^2\to q_{max}^2$) and crosses zero at $q^2\approx8.5$ GeV$^2$.
	
\begin{figure}[!htb]
		\centering
		\includegraphics[width=0.8\linewidth]{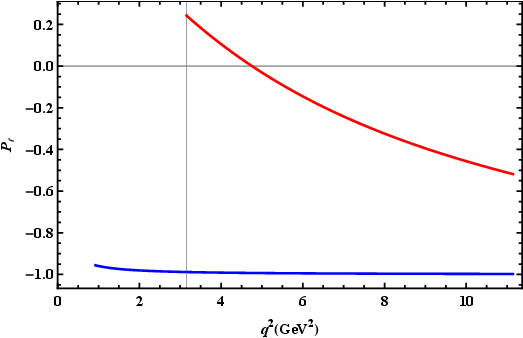}
		\caption{The $q^2$-dependence of the lepton polarization $P_{\ell}$ for the $\mu^{-}$(blue) and $\tau^{-}$(red) modes..}
		\label{fig:PL}
	\end{figure}

\begin{figure}[!htb]
		\centering
		\includegraphics[width=0.8\linewidth]{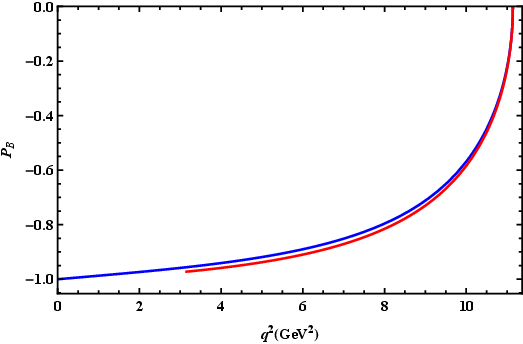}
		\caption{The $q^2$-dependence of the polarization $P_{B}(q^2)$ of the daughter baryon $\Lambda_{c}$ for the $\mu^{-}$(blue) and $\tau^{-}$(red) modes.}
		\label{fig:PB}
	\end{figure}

In Figure.~\ref{fig:PL}, we present the $q^2$-dependence of the lepton polarization asymmetry $P_{\ell}(q^2)$ as defined in Eq. (\ref{eq:PL}). For the $\mu$-mode decays, the blue solid curve corresponds to the chiral limit ($m_\mu \to 0$) where the lepton becomes purely left-handed.  Figure.~\ref{fig:PB} displays the $\Lambda_c$ baryon polarization parameter derived from Eq.(\ref{eq:PB}). For both modes, the hadronic polarization vanishes at zero recoil as a consequence of the helicity amplitude relations  $H_{1/2,1}^{2}-H_{-1/2,-1}^{2}=H_{1/2,0}^{2}-H_{-1/2,0}^{2}=H_{1/2,t}^{2}-H_{-1/2,t}^{2}=0$ at zero recoil.
	
\begin{figure}[!htb]
		\centering
		\includegraphics[width=0.8\linewidth]{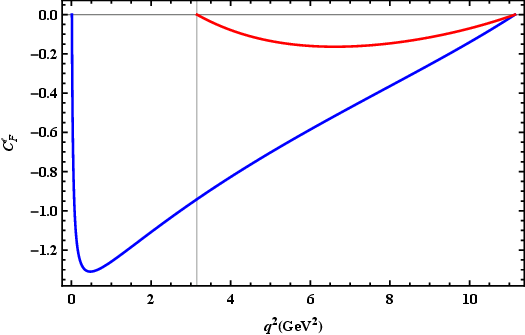}
		\caption{The $q^2$ dependence of the convexity parameter $C_{F}^\ell(q^2)$ for the $\mu^{-}$(blue) and $\tau^{-}$(red) modes.}
		\label{fig:CF}
	\end{figure}

In Figure~\ref{fig:CF}, we show the $q^2$ dependence of the convexity parameter $C_F$, as defined in Eq.(\ref{eq:CF}). At zero recoil, $C_F^\ell$ vanishes for both decay modes, which follows from the relation $H_{1/2,1}^2 + H_{-1/2,-1}^2 = 2(H_{1/2,0}^2 + H_{-1/2,0}^2)$. In the $\tau$ channel, $C_F^\tau$ also approaches zero near the kinematic threshold $q^2 = m_\tau^2$, where the ratio $m_\ell^2/q^2 \to 1$, in accordance with Eq.(\ref{eq:CF}) and Eq.(\ref{eq:dw}). Over the full $q^2$ range accessible in the $\tau$ mode, $C_F^\tau$ remains small, indicating that the $\cos\theta_\ell$ distribution is nearly linear. In contrast, for the $\mu$-mode, $C_F^\mu$ can become significantly negative, corresponding to a strongly parabolic angular distribution characterized by a downward-opening curvature.

\section{Summary}\label{sec:6}
In this work, we calculate the transition form factors for the decay $\Lambda_b \to \Lambda_c$ using the pQCD approach and investigate the semileptonic decay $\Lambda_b \to \Lambda_c \ell^- \bar{\nu}_\ell$ within the SM. Six independent form factors are evaluated in the low-$q^2$ region and then extrapolated to the entire physical kinematic range by parametrizing their $q^2$ dependence using the $z$-expansion, incorporating the latest lattice QCD results in the high-$q^2$ region. We also analyze the $q^2$ behavior of these form factors and find that their magnitudes generally increase with $q^2$, as expected for weak decays. Among them, the dominant contributions arise from the vector and axial-vector form factors $f_1(q^2)$ and $g_1(q^2)$, which exhibit very similar $q^2$ dependence. The remaining form factors, $f_{2,3}(q^2)$ and $g_{2,3}(q^2)$, are found to be negligibly small across the entire $q^2$ range, consistent with expectations from HQET. Our results at both the maximum and minimum recoil points are compared with previous theoretical predictions obtained from various frameworks, including constituent quark models, QCD sum rules, and HQET, showing reasonable agreement.

Utilizing the calculated form factors and helicity formalism, we further investigate several experimentally relevant observables, including branching fractions, forward-backward asymmetries, the lepton-side convexity parameter, and final-state polarizations. Both of the differential and integrated observables across all lepton channels are calculated, which could be tested with future experiments. It is found  that the shape of the differential distributions of these quantities for the $\tau$ mode is different from those for the $\mu$  mode due to the large $\tau$-lepton mass effect. These predictions may confirm  the SM or give complimentary information regarding possible new physics in heavy baryon decays. In addition, we predict the lepton flavor violating ratio $\mathcal{R}_{\Lambda_c}=0.29^{+0.12}_{-0.11}$ that shows good consistency with existing available predictions. However, our result is slightly larger than the latest experimental data \cite{LHCb:2022piu}, $\mathcal{R}_{\Lambda_{c}} = 0.242 \pm 0.026\ (\text{stat}) \pm 0.040\ (\text{syst}) \pm 0.059\ (\text{ext})$, corresponding to a tension of approximately $0.35\sigma$. Although not statistically significant, this mild deviation may hint at a potential violation of LFU in semileptonic $\Lambda_b \to \Lambda_c \ell^- \bar{\nu}_{\ell}$ decays.

We acknowledge that there are considerable theoretical uncertainties in our calculations. In particular, the limited precision of the LCDAs of heavy baryons introduces sizable model dependence in the form factor calculations. Future improvements on this front are essential, including lattice QCD determinations of higher-order LCDA moments and more refined QCD sum rule analyses. Additionally, subleading contributions in the perturbative expansion, such as next-to-leading-order (NLO) corrections in $\alpha_s$, and higher-twist effects beyond the leading-power factorization, must be systematically incorporated to reduce theoretical uncertainties. The inclusion of $1/m_Q$ power corrections within HQET and pQCD frameworks may further enhance the fidelity of the predictions. Such efforts will not only improve the accuracy of $\Lambda_b \to \Lambda_c$ form factors but also increase the reliability of Standard Model benchmarks against which potential new physics effects can be probed.

\section*{Acknowledgments}
We thank Rui Zhou, Fu-sheng Yu, Yue-Long Shen and Jia-Jie Han for useful discussions. This work is supported in part by the National Science Foundation of China under the Grants No. 12375089 and 12435004, and the Natural Science Foundation of Shandong province under the Grant No. ZR2022ZD26 and ZR2022MA035.
	
\begin{appendix}\label{appendix}
\section{The detailed process of calculating form factors using factorizable diagrams.}
\label{sec:appendix}

\begin{table*}[!htb]
\caption{The virtualities of the internal gluon $t_{A,B}^{T_i}$
and quark $t_{C,D}^{T_i}$ for diagram $T_1,T_2,T_3,T_5,T_6,T_7,T_9$ in Fig.\ref{Feynman diagram}.}
\label{tab:hard scale}
\renewcommand{\arraystretch}{1.4}
\setlength{\abovecaptionskip}{2cm}
\setlength{\belowcaptionskip}{-0.5cm}
\setlength{\tabcolsep}{13pt}
\begin{tabular}[t]{ccccc}
\hline\hline
$  T_i$          & $t_{A}^{T_i}$           & $t_{B}^{T_i}$            & $t_{C}^{T_i}$                    & $t_{D}^{T_i}$                      \\ \hline
$T_{1}$  & $m_{\Lambda_{b}}^{2}x_3x_3'$         & $m_{\Lambda_{b}}^{2}(1-x_1)(1-x_1')$   & $m_{\Lambda_{b}}^{2}(1-x_1)x_3'$              & $m_{\Lambda_{b}}^{2}(1-x_1')$                          \\
$T_{2}$  & $m_{\Lambda_{b}}^{2}x_3x_3'$         &$ m_{\Lambda_{b}}^{2}(1-x_1)(1-x_1')$   & $m_{\Lambda_{b}}^{2}(1-x_1)x_3'$              & $m_{\Lambda_{b}}^{2}(1-x_1')$                          \\
$T_{3}$  & $m_{\Lambda_{b}}^{2}x_3x_3'$         & $m_{\Lambda_{b}}^{2}x_2x_2'$         & $m_{\Lambda_{b}}^{2}(x_2+x_3'-x_2x_3')$         & $m_{\Lambda_{b}}^{2}(1-x_1')$                          \\
$T_{5}$  & $m_{\Lambda_{b}}^{2}x_3x_3'$         & $m_{\Lambda_{b}}^{2}x_2x_2'$         & $m_{\Lambda_{b}}^{2}(x_2'+x_3-x_2'x_3)$         & $m_{\Lambda_{b}}^{2}\left[r_c^2+(1-x_2')(x_3-r^2)\right]$         \\
$T_{6}$  & $m_{\Lambda_{b}}^{2}x_3x_3'$         & $m_{\Lambda_{b}}^{2}(1-x_1)(1-x_1')$ & $m_{\Lambda_{b}}^{2}(1-x_1)x_3'$              & $m_{\Lambda_{b}}^{2}\left(1+r_c^2-r^2-x_1\right)$                 \\
$T_{7}$  & $m_{\Lambda_{b}}^{2}x_3x_3'$         & $m_{\Lambda_{b}}^{2}(1-x_1)(1-x_1')$ & $m_{\Lambda_{b}}^{2}(1-x_1')x_3$              & $m_{\Lambda_{b}}^{2}\left(1+r_c^2-r^2-x_1\right)$                 \\
$T_{9}$  & $m_{\Lambda_{b}}^{2}x_3x_3'$         & $m_{\Lambda_{b}}^{2}x_2x_2'$         & $m_{\Lambda_{b}}^{2}\left(1+r_c^2-r^2-x_1\right)$          & $m_{\Lambda_{b}}^{2}\left[r_c^2+x_2+r^2(x_3'-1)-x_2 x_3'\right]$   \\
\hline\hline
\end{tabular}
\end{table*}
\begin{table*}[!htb]
\caption{The expressions of $B_{T_{i}}$ for diagrams $T_1,T_2,T_3,T_5,T_6,T_7,T_9$ in Fig.\ref{Feynman diagram}.}
\label{tab:Db}
\setlength{\tabcolsep}{24pt}
\begin{tabular}{cc}
\hline\hline
$T_i$          &$B_{T_i}$\\ \hline
$T_1$  & $\frac{1}{(2\pi)^4}K_0(\sqrt{t_A}|\textbf{b}_2-\textbf{b}_3|)K_0(\sqrt{t_B}|\textbf{b}_3+\textbf{b}'_3-\textbf{b}'_2|)
 K_0(\sqrt{t_C}|\textbf{b}_2-\textbf{b}_3+\textbf{b}'_2-\textbf{b}'_3|)K_0(\sqrt{t_D}|\textbf{b}_3+\textbf{b}'_3|)$ \\

 $T_2$ & $\frac{1}{(2\pi)^4}K_0(\sqrt{t_A}|\textbf{b}_2-\textbf{b}_3|)K_0(\sqrt{t_B}|\textbf{b}_2|)
 K_0(\sqrt{t_C}|\textbf{b}'_2-\textbf{b}'_3+\textbf{b}_2-\textbf{b}_3|)K_0(\sqrt{t_D}|\textbf{b}_3+\textbf{b}'_3|)$ \\

$T_3$  &$\frac{1}{\left(2\pi\right)^5}K_0(\sqrt{t_A}|\textbf{b}_3|)h_1$ \\

$T_5$  &$\frac{1}{\left(2\pi\right)^4}K_0(\sqrt{t_A}|\textbf{b}'_3|)K_0(\sqrt{t_B}|\textbf{b}_2|)h_2$ \\

$T_6$  & $\frac{1}{(2\pi)^5}K_0(\sqrt{t_A}|\textbf{b}_2-\textbf{b}_3|)K_0(\sqrt{t_B}|\textbf{b}'_2|)
 K_0(\sqrt{t_C}|\textbf{b}_2-\textbf{b}_3+\textbf{b}'_2-\textbf{b}'_3|)h_3$ \\

$T_7$ & $\frac{1}{(2\pi)^5}K_0(\sqrt{t_A}|\textbf{b}'_2-\textbf{b}'_3|)K_0(\sqrt{t_B}|\textbf{b}_2-\textbf{b}_3-\textbf{b}'_3|)
 K_0(\sqrt{t_C}|\textbf{b}_2-\textbf{b}_3+\textbf{b}'_2-\textbf{b}'_3|)h_4$ \\

$T_9$  &$\frac{1}{\left(2\pi\right)^5}K_0(\sqrt{t_B}|\textbf{b}'_2|)h_5$ \\

\hline\hline
\end{tabular}
\end{table*}

The virtualities of two internal gluons
and two quarks for each diagram $T_{i}$ in Fig.\ref{Feynman diagram} are collected in Table~\ref{tab:hard scale}. The inner functions $B_{T_i}$ from the denominators of the internal propagators in diagrams $T_i$ are presented in Table.~\ref{tab:Db}, where the functions $h_l$ are defined as
\begin{widetext}
\begin{eqnarray}
h_{1}&=&\int_{0}^{1}\frac{dz_{1}dz_{2}}{z_{1}\left(1-z_{1}\right)}\frac{\sqrt{X_{1}}}{\sqrt{\left|Y_{1}\right|}}\pi^2 K_{1}\left(\sqrt{X_{1}Y_{1}}\right),\\
h_2&=&\int_{0}^{1}dz\frac{\left|\textbf{b}_{2}+\textbf{b}_{2}^{\prime}\right|}{\sqrt{\left|X_{2}\right|}}\left\{\pi K_{1}\left(\sqrt{X_{2}}\left|\textbf{b}_{2}+\textbf{b}_{2}^{\prime}\right|\right)\theta(X_2)+\frac{\pi^2}{2}\left[N_{1}\left(\sqrt{\left|X_{2}\right|}\left|\textbf{b}_{2}+\textbf{b}_{2}^{\prime}\right|\right)-iJ\left(\sqrt{\left|X_{2}\right|}\left|\textbf{b}_{2}+\textbf{b}_{2}^{\prime}\right|\right)\right]\theta\left(-X_{2}\right)\right\},\\
h_{3}&=&2\pi K_{0}\left(\sqrt{t_{D}^{T_6}}\left|\textbf{b}_{3}+\textbf{b}_{3}^{\prime}\right|\right)\theta\left(t_{D}^{T_6}\right)+\pi^2\left[-N_{0}\left(\sqrt{|t_{D}^{T_6}|}\left|\textbf{b}_{3}+\textbf{b}_{3}^{\prime}\right|\right)+iJ_{0}\left(\sqrt{|t_{D}^{T_6}|}\left|\textbf{b}_{3}+\textbf{b}_{3}^{\prime}\right|\right)\right]\theta\left(-t_{D}^{T_6}\right).\\
h_{4}&=&2\pi K_{0}\left(\sqrt{t_{D}^{T_7}}\left|\textbf{b}_{3}+\textbf{b}_{3}^{\prime}\right|\right)\theta\left(t_{D}^{T_7}\right)+\pi^2\left[-N_{0}\left(\sqrt{|t_{D}^{T_7}|}\left|\textbf{b}_{3}+\textbf{b}_{3}^{\prime}\right|\right)+iJ_{0}\left(\sqrt{|t_{D}^{T_7}|}\left|\textbf{b}_{3}+\textbf{b}_{3}^{\prime}\right|\right)\right]\theta\left(-t_{D}^{T_{7}}\right).\\
h_{5}&=&\int_{0}^{1}\frac{dz_{1}dz_{2}}{z_{1}\left(1-z_{1}\right)}\frac{\sqrt{X_{3}}}{\sqrt{\left|Y_{3}\right|}}\left\{\pi^2 K_{1}\left(\sqrt{X_{3}Y_{3}}\right)\theta\left(Y_{3}\right)+\frac{\pi^3}{2}\left[N_{1}\left(\sqrt{X_{3}\left|Y_{3}\right|}\right)-iJ_{1}\left(\sqrt{X_{3}\left|Y_{3}\right|}\right)\right]\theta\left(-Y_{3}\right)\right\},
\end{eqnarray}
with
\begin{eqnarray}
X_{1}&=&\left|\left(1-z_{1}\right)\textbf{b}_{2}^{\prime}+z_{1}\textbf{b}_{2}\right|^{2}+\frac{z_{1}\left(1-z_{1}\right)}{z_{2}}\left|\textbf{b}_{2}+\textbf{b}_{2}^{\prime}\right|^2,
\notag\\
Y_{1}&=&t_{B}^{T_3}\left(1-z_{2}\right)+\frac{z_{2}}{z_{1}\left(1-z_{1}\right)}\left\{t_{C}^{T_3}\left(1-z_{1}\right)+t_{D}^{T_3}z_{1}\right\}.\\
X_{2}&=&t_{C}^{T_5}z+t_{D}^{T_5}\left(1-z\right).\\
X_{3}&=&\left|-\textbf{b}_{2}^{\prime}-\textbf{b}_{2}+\textbf{b}_{3}^{\prime}+z_{1}\left(\textbf{b}_{2}+\textbf{b}_{2}^{\prime}\right)\right|^{2}+\frac{z_{1}\left(1-z_{1}\right)}{z_{2}}\left|\textbf{b}_{2}+\textbf{b}_{2}^{\prime}\right|^2,\notag\\
Y_{3}&=&t_{A}^{T_9}\left(1-z_{2}\right)+\frac{z_{2}}{z_{1}\left(1-z_{1}\right)}\left[t_{C}^{T_9}\left(1-z_{1}\right)+t_{D}^{T_9}z_{1}\right].
\end{eqnarray}
\end{widetext}
with the Bessel functions $K_{0,1}$, $N_{0,1}$, and $J_1$ and $\theta(x)$ is a $\theta$ function. Below we provide the Fourier integration formula and various Bessel functions used in calculating the form factor. Among them, the Bessel function $K_{n}$ has such a relationship with the first type Bessel function $J_{n}$ and the second type Bessel function $N_{n}$
\begin{widetext}
\begin{eqnarray}
&&K_{n}\left(-iz\right)=\frac{i\pi}{2}e^{(in\pi)/2}\left[J_{n}(z)+iN_{n}(z)\right]\\
&&\int d^{2}k_{T}\frac{e^{i\textbf{k}_{T}\cdot\textbf{b}}}{k_{T}^{2}+A}=2\pi K_{0}\left(\sqrt{A\left|\textbf{b}\right|}\right)\theta(A)+\pi^2\left[-N_{0}\left(\sqrt{\left|A\right|}\left|\textbf{b}\right|\right)+iJ_{0}\left(\sqrt{\left|A\right|}\left|\textbf{b}\right|\right)\right]\theta(-A),
\\
&&\int d^{2}k_{T}\frac{e^{i\textbf{k}_{T}\cdot \textbf{b}}}{\left(k^2_{T}+A\right)\left(k^2_{T}+B\right)}=\int_{0}^{1}dz\frac{\left|\textbf{b}\right|}{\sqrt{\left|\textbf{Z}_1\right|}}\Big\{\pi K_{1}\left(\sqrt{Z_1}\left|\textbf{b}\right|\right)\theta\left(Z_1\right)+\frac{\pi^2}{2}\left[N_{1}\left(\sqrt{\left|Z_1\right|}\left|\textbf{b}\right|\right)-iJ_{1}\left(\sqrt{\left|Z_1\right|}\left|\textbf{b}\right|\right)\right]\theta\left(-Z_1\right)\Big\},\notag\\
\\
&&\int d^{2}k_{1T}d^{2}k_{2T}\frac{e^{i\textbf{k}_{1T}\cdot\textbf{b}_{1}+\textbf{k}_{2T}\cdot\textbf{b}_{2}}}{\left(k_{1T}^{2}+A\right)\left(k_{2T}^{2}+B\right)\left[\left(k_{1T}+k_{2T}\right)^2+C\right]}=\int_{0}^{1}\frac{dz_{1}dz_{2}}{z_{1}\left(1-z_{1}\right)}\frac{\sqrt{Y}}{\sqrt{\left|Z_{2}\right|}}\Big\{\pi^2 K_{1}\left(\sqrt{YZ_{2}}\right)\theta\left(Z_{2}\right)\notag\\
&&\hspace{8cm}+\frac{\pi^3}{2}\left[N_{1}\left(\sqrt{Y|Z_{2}|}\right)-iJ_{1}\left(\sqrt{Y|Z_{2}|}\right)\right]\theta\left(-Z_{2}\right)\Big\},
\end{eqnarray}
\end{widetext}
and the parameters $Z_{1},Z_{2}$ and $Y$ are given by
\begin{align}
Z_{1}=&zA+\left(1-z\right)B,\\
Z_{2}=&\left(1-z_{2}\right)A+\frac{z_2}{z_1\left(1-z_1\right)}\left[\left(1-z_1\right)B+z_1C\right],\\
Y=&\left(\textbf{b}_1-z_{1}\textbf{b}_2\right)^2+\frac{z_1\left(1-z_1\right)}{z_2}\textbf{b}_{2}^2.
\end{align}

The formulas of the hard scatter function $H_k(x_i,x^\prime_i)$ from the Feynman diagrams are as below
\begin{widetext}
\begin{align}
H_{f_1}^{T_1}=&H_{g_1}^{T_1}=8m_{\Lambda_b}^2\left\{x_{3}^{\prime}\left(2r-1\right)-\left(x_1-1\right)x_1^{\prime}\left(r-2\right)r\right\},\\
H_{f_2}^{T_1}=&H_{f_3}^{T_1}=\frac{8m_{\Lambda_b}^4}{\left(r^2-1\right)}\left\{\left(x_{1}^{\prime}-1\right)\left(1-x_1\right)\left(r-2\right)r\right\},  \\
H_{g_2}^{T_1}=&H_{g_3}^{T_1}=\frac{8m_{\Lambda_b}^4}{\left(r^2-1\right)}\left\{\left(x_{1}^{\prime}-1\right)\left(x_1-1\right)\left(r-2\right)r\right\},\\
H_{f_{1}}^{T_2}=&H_{g_{1}}^{T_2}=-8m_{\Lambda_b}^4\left\{x_1^{\prime}\left(r^2-2\left(x_3+1\right)r+x_3\right)-\left(r-2\right)r\right\},\\
H_{f_{2}}^{T_2}=&H_{f_{3}}^{T_2}=\frac{8m_{\Lambda_b}^4}{\left(r^2-1\right)}\left\{\left(x_1^{\prime}-1\right)x_3\left(2r-1\right)\right\},\\
H_{g_{2}}^{T_2}=&H_{g_{3}}^{T_2}=-\frac{8m_{\Lambda_b}^4}{\left(r^2-1\right)}\left\{\left(x_1^{\prime}-1\right)x_3\left(2r-1\right)\right\},\\
H_{f_1}^{T_3}=&H_{g_1}^{T_3}=4m_{\Lambda_{b}
}^4\left\{r\left(2x_1^{\prime}r+r-2\right)x_3^{\prime}+x_3^{\prime}+x_2\left(x_1^{\prime}\left(r-1\right)^2+2\right)-2\left(r+1\right)\left(x_1^{\prime}r+1\right)\right\},\\
H_{f_2}^{T_3}=&H_{f_3}^{T_3}=\frac{4m_{\Lambda_b}^4}{r^2-1}\left(x_1^{\prime}-1\right)\Big\{r^2\left(x_2+2x_3^{\prime}-2\right)-2\left(x_2+1\right)r+x_2\Big\},\\
H_{g_2}^{T_3}=&H_{g_3}^{T_3}=-\frac{4m_{\Lambda_b}^4}{r^2-1}\left(x_1^{\prime}-1\right)\Big\{r^2\left(x_2+2x_3^{\prime}-2\right)-2\left(x_2+1\right)r+x_2\Big\},\\
H_{f_1}^{T_5}=&-4m_{\Lambda_b}^4\Big\{\left(x_2^{\prime}-1\right)x_2^{\prime}r^3+r^2\left(2\left(r_c+1\right)-x_2^{\prime})\left(r_c+2\right)\right)+\left(x_2^{\prime}-1\right)\left(x_2^{\prime}-2\left(r_c+1\right)\right)r+x_2^{\prime}r_c\Big\},\\
H_{f_{2}}^{T_5}=&\frac{4m_{\Lambda_b}^4}{\left(r^2-1\right)^2}\Big\{\left(r-1\right)r\left(r^2+1\right)x_2^{\prime2}-r_cx_2^{\prime}+2r\left(-r^2+x_3+1\right)\left(r-r_c\right)\notag\\
&+r\left(-3r+\left(r-1\right)\left(r\left(r-r_c\right)-2x_3\right)+3r_c+1\right)x_2^{\prime}\Big\},\\
H_{f_3}^{T_5}=&\frac{-4m_{\Lambda_{b}}^4}{\left(1-r^2\right)^2}\left(r-1\right)\Big\{-2\left(r^2+1\right)x_3^{2}+2r\left(\left(r+1\right)\left(r^2-r_cr+2r_c+2\right)-x_2^{\prime}\left(r+1\right)^2\right)x_3+x_2^{\prime2}r\left(r^2+1\right)^2\notag\\
&+2\left(r-1\right)r\left(r+1\right)^2\left(r_c-r\right)+x_2^{\prime}\left(r+1\right)\left(r_c+r\left(-3r_c+r\left(r\left(r\left(r-r_c-5\right)+r_c-1\right)-1\right)-1\right)\right)\Big\},\\
H_{g_1}^{T_5}=&4m_{\Lambda_b}^4\Big\{\left(x_2^{\prime}-1\right)r\left(\left(r-4\right)rx_2^{\prime}+x_2^{\prime}+2\left(r+1\right)\right)+\left(r\left(3r-2\right)x_2^{\prime}+x_2^{\prime}-2r\left(r+1\right)\right)r_c\Big\},\\
H_{g_2}^{T_5}=&\frac{-4m_{\Lambda_b}^4}{\left(r^2-1\right)}\Big\{r\left(r+1\right)\left(\left(r-4\right)r+1\right)x_2^{\prime2}+\left(r_c-r\left(-2x_3\left(r+1\right)+r_c+r\left(r\left(3r-3r_c-5\right)+r_c-5\right)+1\right)\right)x_2^{\prime}\notag\\
&+2r\left(r^2-1\right)\left(r-r_c\right)-2x_3r\left(r+r_c+2\right)\Big\},\\
H_{g_{3}}^{T_5}=&\frac{4m_{\Lambda_{b}}^4}{\left(1-r^2\right)^2}\left(r+1\right)\Big\{r\left(\left(r-4\right)r+1\right)\left(r^2+1\right)x_2^{\prime2}-\left(r-1\right)\left(3r^3-r^2+r-1\right)\left(r-r_c\right)x_2^{\prime}+2x_3^{2}\left(\left(r-4\right)r+1\right)\notag\\
&+2x_3r\left(x_2^{\prime}\left(r+1\right)^2-\left(r-2\right)\left(r-1\right)\left(r-r_c\right)\right)+2\left(r-1\right)^2r\left(r+1\right)\left(r-r_c\right)\Big\}.\\
H_{f_1}^{T_6}=&H_{g_1}^{T_6}=-8m_{\Lambda_{b}}^4r\Big\{x_1\left(r-2\right)+x_3^{\prime}r_c-r\left(2x_3^{\prime}r_c+1\right)+2\Big\},\\
H_{f_2}^{T_6}=&H_{g_2}^{T_6}=\frac{8m_{\Lambda_b}^4r}{\left(r^2-1\right)^2}\Big\{\left(x_1-1\right)\left(r-2\right)\left(r-r_c\right)+x_3^{\prime}\left(2r-1\right)\left(r^2-r_cr+x_1-1\right)\Big\},\\
H_{f_3}^{T_6}=&H_{g_3}^{T_6}=-\frac{8m_{\Lambda_b}^4r}{\left(r^2-1\right)^2}\Big\{\left(x_1-1\right)\left(r-2\right)\left(r-r_c\right)+x_3^{\prime}\left(2r-1\right)\left(r^2-r_cr+x_1-1\right)\Big\},\\
H_{f_1}^{T_7}=&H_{g_1}^{T_7}=-8m_{\Lambda_b}^4\Big\{\left(x_1^{\prime}-1\right)\left(r-2\right)r_cr^2-2x_3r+x_3\Big\},\\
H_{f_2}^{T_7}=&H_{g_2}^{T_7}=-\frac{8m_{\Lambda_b}^4}{\left(r^2-1\right)}\Big\{\left(x_1^{\prime}-1\right)\left(r-2\right)\left(r^2-r_cr+x_1-1\right)r^2+x_3\left(2r-1\right)\left(r-r_c\right)\Big\},\\
H_{f_3}^{T_7}=&H_{g_3}^{T_7}=\frac{8m_{\Lambda_b}^4}{\left(r^2-1\right)}\Big\{\left(x_1^{\prime}-1\right)\left(r-2\right)\left(r^2-r_cr+x_1-1\right)r^2+x_3\left(2r-1\right)\left(r-r_c\right)\Big\},\\
H_{f_1}^{T_9}=&H_{g_1}^{T_9}=-4m_{\Lambda_{b}}^4\Big\{\left(-2x_{3}^{\prime}+\left(r-2\right)r+2\right)r^2-\left(\left(x_{3}^{\prime}-2\right)r^2-2\left(x_{3}^{\prime}+1\right)r+x_{3}^{\prime}+2\right)r_{c}r+x_{3}\left(r+2r_{c}-2\right)r+2r\notag\\
&-\left(\left(r-4\right)r+1\right)r_{c}^{2}+x_3+x_1\left(\left(r-1\right)^2+2rr_c\right)-1\Big\}\\
H_{f_2}^{T_9}=&H_{g_2}^{T_9}=\frac{4m_{\Lambda_b}^{4}}{r^2-1}\Big\{\left(x_{3}^{\prime}-2\right)r^5-\left(x_3^{\prime}\left(r_c+2\right)-2r_c\right)r^4+\left(-x_3+x_1\left(x_3^{\prime}-4\right)+2x_3^{\prime}\left(r_c-1\right)+6\right)r^3\notag\\
&+\big(\big(2x_1+x_3+x_3^{\prime}-4\big)r_c-2\left(x_3+\left(x_1-1\right)x_3^{\prime}\right)\big)r^2+\big(-2x_1^2+\left(-2x_3+x_3^{\prime}-2r_c+4\right)x_1+3x_3-x_3^{\prime}\notag\\
&+2\left(x_3+1\right)r_c-2\big)r-x_3r_c\Big\}\\
H_{f_3}^{T_9}=&H_{g_3}^{T_9}=\frac{4m_{\Lambda_{b}}^4}{r^2-1}\Big\{-\left(\left(x_3^{\prime}-2\right)r^5\right)+\left(x_3^{\prime}\left(r_{c}+2\right)-2r_c\right)r^4+\left(x_3-x_1\left(x_3^{\prime}-4\right)+2x_3^{\prime}-2x_3^{\prime}r_c-6\right)r^3\notag\\
&+\big(2\left(x_3+\left(x_1-1\right)x_3^{\prime}\right)-\left(2x_1+x_3+x_3^{\prime}-4\right)r_c\big)r^2\notag\\
&+\left(-3x_3+2x_1\left(x_1+x_3-2\right)-x_1x_3^{\prime}+x_3^{\prime}+2\left(x_1-x_3\right)r_c-2r_c+2\right)r+x_3r_c\Big\}
\end{align}
\end{widetext}
\end{appendix}		
\bibliographystyle{bibstyle}
\bibliography{reference}
\end{document}